\documentclass[conference]{IEEEtran}
\usepackage{amsmath,amssymb,amsfonts}
\usepackage{algorithmic}
\usepackage{graphicx}
\usepackage{textcomp}
\usepackage{xcolor}
\usepackage{subfigure}
\usepackage{comment}
\usepackage{adjustbox}
\usepackage{multirow,tabularx}
\usepackage{array,etoolbox}
\usepackage{enumitem}
\preto\tabular{\setcounter{magicrownumbers}{0}}
\newcounter{magicrownumbers}

\usepackage{mathptmx} % This is Times font

\usepackage{fancyhdr}
\usepackage[normalem]{ulem}
\usepackage[hyphens]{url}
\usepackage[sort,nocompress]{cite}
\usepackage[final]{microtype}
\usepackage[keeplastbox]{flushend}
\usepackage[bookmarks=true,breaklinks=true,letterpaper=true,colorlinks,citecolor=blue,linkcolor=blue,urlcolor=blue]{hyperref}

\usepackage{xcolor,colortbl}
\definecolor{codegreen}{rgb}{0,0.6,0}
\definecolor{codegray}{rgb}{0.5,0.5,0.5}
\definecolor{codepurple}{rgb}{0.58,0,0.82}
\definecolor{backcolour}{rgb}{0.95,0.95,0.92}
\definecolor{LightBlue}{rgb}{0.83, 0.91, 1}
\definecolor{LightGreen}{rgb}{0.8, 1, 0.8}
\definecolor{LightPink}{rgb}{1, 0.8, 0.88}
\definecolor{LightYellow}{rgb}{1, 1, 0.6}
\definecolor{PaleYellow}{rgb}{1, 0.98, 0.9}
\definecolor{light-gray}{gray}{0.9}

\newcolumntype{P}[1]{>{\centering\arraybackslash}p{#1}}
\newcommand{\wname}{ULEEN}
\definecolor{revhl}{RGB}{200,0,20}
\newcommand{\revision}[1]{{\textcolor{revhl}{#1}}}
\renewcommand{\revision}[1]{{#1}} % Uncommenting this disables reformatting of revision blocks

\def\BibTeX{{\rm B\kern-.05em{\sc i\kern-.025em b}\kern-.08em
    T\kern-.1667em\lower.7ex\hbox{E}\kern-.125emX}}
\begin{document}

\title{\wname: A Novel Architecture for Ultra Low-Energy Edge Neural Networks}

\author{
\IEEEauthorblockN{Zachary Susskind, 
Aman Arora}
\IEEEauthorblockA{
\textit{The University of Texas at Austin}
}
\IEEEauthorblockN{
Igor Dantas Dos Santos Miranda}
\IEEEauthorblockA{
\textit{Federal University of Recôncavo da Bahia}
}
\IEEEauthorblockN{
Alan Tendler Leibel Bacellar, Luis Armando Quintanilla Villon,
Rafael Fontella Katopodis}
\IEEEauthorblockA{
\textit{Federal University of Rio de Janeiro}
}
\IEEEauthorblockN{
Leandro Santiago de Araújo}
\IEEEauthorblockA{
\textit{Universidade Federal Fluminense}
}
\IEEEauthorblockN{
Diego Leonel Cadette Dutra,
Priscila Machado Vieira Lima}
\IEEEauthorblockA{
\textit{Federal University of Rio de Janeiro}
}
\IEEEauthorblockN{
Felipe Maia Galvão França}
\IEEEauthorblockA{
\textit{Instituto de Telecomunicações, Porto}\\
\textit{Federal University of Rio de Janeiro}
}
\IEEEauthorblockN{
Mauricio Breternitz Jr.}
\IEEEauthorblockA{
\textit{ISCTE - Instituto Universitario de Lisboa}
}
\IEEEauthorblockN{
Lizy K. John
}
\IEEEauthorblockA{
\textit{The University of Texas at Austin}
}
\vspace{-2ex}
}

%\author{\IEEEauthorblockN{1\textsuperscript{st} Given Name Surname}
%\IEEEauthorblockA{\textit{dept. name of organization (of Aff.)} \\
%\textit{name of organization (of Aff.)}\\
%City, Country \\
%email address or ORCID}
%\and
%\IEEEauthorblockN{2\textsuperscript{nd} Given Name Surname}
%\IEEEauthorblockA{\textit{dept. name of organization (of Aff.)} \\
%\textit{name of organization (of Aff.)}\\
%City, Country \\
%email address or ORCID}
%\and
%\IEEEauthorblockN{3\textsuperscript{rd} Given Name Surname}
%\IEEEauthorblockA{\textit{dept. name of organization (of Aff.)} \\
%\textit{name of organization (of Aff.)}\\
%City, Country \\
%email address or ORCID}
%\and
%\IEEEauthorblockN{4\textsuperscript{th} Given Name Surname}
%\IEEEauthorblockA{\textit{dept. name of organization (of Aff.)} \\
%\textit{name of organization (of Aff.)}\\
%City, Country \\
%email address or ORCID}
%\and
%\IEEEauthorblockN{5\textsuperscript{th} Given Name Surname}
%\IEEEauthorblockA{\textit{dept. name of organization (of Aff.)} \\
%\textit{name of organization (of Aff.)}\\
%City, Country \\
%email address or ORCID}
%\and
%\IEEEauthorblockN{6\textsuperscript{th} Given Name Surname}
%\IEEEauthorblockA{\textit{dept. name of organization (of Aff.)} \\
%\textit{name of organization (of Aff.)}\\
%City, Country \\
%email address or ORCID}
%}

\maketitle
\pagestyle{plain}

\begin{abstract}
The deployment of AI models on low-power, real-time edge devices requires accelerators for which energy, latency, and area are all first-order concerns. 
There are many approaches to enabling deep neural networks (DNNs) in this domain, including pruning, quantization, compression, and binary neural networks (BNNs), but with the emergence of the ``extreme edge'', there is now a demand for even more efficient models. 
%In order to meet the constraints of this new domain, we need to look beyond DNNs into fundamentally different approaches to AI.
In order to meet the constraints of ultra-low-energy devices, we propose \wname, a model architecture based on weightless neural networks.

Weightless neural networks (WNNs) are a class of neural model which use table lookups, not arithmetic, to perform computation. The elimination of energy-intensive arithmetic operations makes WNNs theoretically well suited for edge inference; however, they have historically suffered from poor accuracy and excessive memory usage.
%In this paper, we first propose
\wname~ incorporates algorithmic improvements and a novel training strategy inspired by BNNs to make significant strides in improving accuracy and reducing model size. %Compared to a state-of-the-art WNN model architecture across nine classification datasets, \wname~reduces mean model size by 46.1\% and test error by 49.8\%.
%on average reduces error by more than 40\% with a more than 50\% reduction in model size.
%{\color{red}In this paper, we propose \wname, a novel neural network model architecture which incorporates principles from weightless neural networks, binary neural network training, and a variety of algorithmic and architectural improvements to achieve a new state of the art edge neural network surpassing prior BNNs, WNNs and quantized DNNs in energy efficiency and latency.}

We compare FPGA and ASIC implementations of an inference accelerator for \wname~against edge-optimized DNN and BNN devices. On a Xilinx Zynq Z-7045 FPGA, we demonstrate classification on the MNIST dataset at 14.3 million inferences per second (13 million inferences/Joule) with 0.21 $\mu$s latency and 96.2\% accuracy, while Xilinx FINN achieves 12.3 million inferences per second (1.69 million inferences/Joule) with 0.31 $\mu$s latency and 95.83\% accuracy. %Considering energy, \wname~ makes 13 million inferences/Joule while FINN makes 1.69 million inferences/Joule.
%an average 2.0x reduction in latency, 2.3x increase in throughput, and 7.7-8.2x improvement in energy per inference versus the Xilinx FINN accelerator for BNN inference. 
In a 45nm ASIC, we achieve 5.1 million inferences/Joule and 38.5 million inferences/second at 98.46\% accuracy, while a quantized Bit Fusion model achieves 9230 inferences/Joule and 19,100 inferences/second at 99.35\% accuracy. In our search for ever more efficient edge devices, \wname~shows that WNNs are deserving of consideration.
%In the search for ever more efficient edge devices, these results show the utility of the \wname~approach which uses WNNs.
% We are the first to achieve... - is there anything to put in the abstract here?

%We then demonstrate the viability of the \wname~architecture for ultra-low-energy inference by comparing FPGA and ASIC prototypes against several edge-optimized DNN devices. On an XX Zynq Z7045 FPGA, we demonstrate classification on the MNIST dataset at XX 14.3 million frames per second with XX ns latency and XX 96.17 accuracy, surpassing the FINN accelerator by  XX improvement in latency, XX improvement in throughput, and XX improvement in energy versus the FINN accelerator for binary neural networks. In our XX 45nm ASIC, we achieve 5 million  inferences/Joule at 98.4\% accuracy whereas bit-fusion achieves 9000 inferences/Joule at 99.35\% accuracy. XX better latency and XX to YY better energy compared to the BitFusion low-precision architecture.
\end{abstract}

%\begin{IEEEkeywords}
%weightless neural networks, WNN, WiSARD, neural networks
%\end{IEEEkeywords}

\section{Introduction}
\label{sec:introduction}

In the last decade, deep neural networks (DNNs) have driven revolutionary improvements in domains including image classification, natural language processing (NLP), and medical diagnostics, in some cases achieving superhuman accuracy~\cite{russa2015}. This advancement has been driven by an exponential increase in the size and complexity of the models we can train. At the same time, there is a second movement in the opposite direction: deployment of AI models on \textit{smaller and smaller devices}. Techniques such as pruning, compression, low-precision quantization~\cite{gholami2021survey}, and Once-for-All~\cite{once_for_all} can greatly reduce the memory requirements and computational demands of large models, enabling their use on low-power, resource-constrained edge devices. However, on the ``extreme edge'', where computation is performed directly adjacent to physical sensors~\cite{extreme_edge_sensors}, energy budgets are minuscule and efficiency is of utmost importance. Such devices are often deployed in inaccessible environments, and may be expected to last years or decades on a small battery, or by harvesting energy directly from the phenomena they measure~\cite{extreme_edge_overview}. 

Binary neural networks (BNNs)~\cite{binaryconnect, binary_net, finn, xnor_net, courbariaux2016binarized} take quantization to its logical extreme, reducing network weights and activations to single-bit values. By replacing energy-intensive multiplication with XNOR operations, they achieve energy efficiency orders of magnitude better than DNNs. However, like DNNs, BNNs must propagate activations through many layers of computation. This may still present a significant critical path for real-time edge applications.

\begin{comment}
\begin{table}[htbp]
\centering
\caption{Weights and MACs for popular DNNs~\cite{vivienne17, mythic}.}
\makebox[\columnwidth][c]{\resizebox{1.1\columnwidth}{!}{
\begin{tabular}{|c|c|c|c|c|c|} 
 \hline
 \rowcolor{LightBlue} \centering
 Metric & LeNet-5 & AlexNet & VGG-16 & Resnet-50 & OpenPose\\
 \hline
 \#Weights & 60k & 61M & 138M &  25.5M & 46M \\
 \hline
 \#MACs & 341k & 724M & 15.5G & 3.9G & 180G\\
 \hline
 Year & 1998 & 2012 & 2014 & 2015 & 2018\\
 \hline
\end{tabular}
}}
\label{tab:dnnsize}
\end{table}
\end{comment}

Weightless Neural Networks (WNNs) are a distinct class of neural model which perform computation primarily using lookup tables, rather than arithmetic or logical functions. Individual weightless neurons, also known as RAM nodes, concatenate binary inputs to form an address into their lookup table, and produce a binary response. WNNs are inspired by the dendritic trees of biological neurons, which perform highly nonlinear decode processing~\cite{wnn_intro_esann}. Unlike traditional or binary neurons, RAM nodes are capable of learning \textit{nonlinear} functions of their inputs.

The concept of WNNs is not new - the earliest implementations date to the 1950s~\cite{weightless_review}. Their simple structure was well-suited to early VLSI techniques, and they achieved some prominence in the 1970s-80s. However, these early WNNs were surpassed by DNNs in both accuracy and memory efficiency by the 1990s. Thus, while the simplicity and non-linearity of WNNs makes them appealing for deployment on the edge, algorithmic and architectural enhancements are needed to make them comparable or superior to optimized DNN and BNN models.

In this paper, we demonstrate techniques to improve the accuracy and reduce the hardware requirements of WNNs.
%These techniques broadly fall into two categories. First, we incorporate improved versions of techniques used separately in prior works but not previously in the same model. 
We present a weightless neural architecture we call \wname~- \textbf{U}ltra \textbf{L}ow \textbf{E}nergy \textbf{E}dge \textbf{N}etworks, suitable for inference applications under extreme energy constraints.
\wname~ incorporates  a set of techniques not previously used in WNNs, including efficient ensembles, pruning, and a multi-epoch gradient-based training algorithm using the straight-through estimator~\cite{straight_through_estimator}.
We also improve upon prior work by including counting Bloom filters with hardware-friendly hashing, a novel nonlinear thermometer encoding, and threshold-based one-shot training~\cite{Grieco:2010:PPE:1751674.1751890, bleaching}.
We present FPGA and ASIC implementations of an inference accelerator for this architecture and compare it against state-of-the-art efficient inference platforms for DNNs and BNNs.

Our specific contributions in this paper are as follows:
\begin{enumerate}
    \item \wname, a weightless neural architecture which incorporates principles from prior WNNs, BNN-inspired training methodologies, and further algorithmic enhancements. Through these improvements, we increase the best reported WNN accuracy on the MNIST dataset
    %~\cite{lecun-mnisthandwrittendigit-2010}
    from 91.5\% to 98.46\%.
    \item An energy-efficient inference accelerator architecture for \wname~which can be implemented on an FPGA or as an ASIC.
    \item Comparisons of our \wname~accelerator with state-of-the-art quantized, optimized DNNs and BNNs on ASIC and FPGA. In comparison to FINN~\cite{finn} (FPGA), we demonstrate 1.4-2.6x improved latency, 1.2-2.6x improved throughput, and 6.8-8.5x improved energy at equal or better accuracy. In comparison to Bit Fusion~\cite{bitfusion:isca:2018} (ASIC), we demonstrate 2014-19549x improved throughput and 479-663x improved energy with <1\% reduction in accuracy. 
    %\item Comparison of \wname~models to a state-of-the-art WNN across nine datasets, with a mean 46.1\% reduction in model size and 49.8\% reduction in error.
    \item A toolchain for generating \wname~models using either single- or multi-shot training rules. A second toolchain which produces RTL from trained \wname~models using our accelerator architecture. These are available at: \textit{URL omitted for double-blinding}.
    
\end{enumerate}
%\end{comment}

The remainder of our paper is organized as follows: In Section \ref{sec:background}, we provide additional background on WNNs and one specific WNN, WiSARD, which we use as the base model for our improvements. In Section \ref{sec:proposal}, we present the \wname~architecture in detail. In Section \ref{sec:methodology}, we discuss software and hardware implementation details. In Section \ref{sec:results}, we compare our accelerator architecture against prior DNN and BNN accelerators and our model architecture against prior memory-efficient WNNs, and provide additional results exploring the performance of our models. In Section \ref{sec:related_work}, we give some additional context into the prior work in this domain. Lastly, in Section \ref{sec:conclusion}, we discuss future work and conclude.
 
\section{Background}
\label{sec:background}

%\subsection{Weightless Neural Networks}
{\bf Weightless neural networks (WNNs)} are neural models which perform computation using lookup tables. The basic computational unit, the RAM node, is an $n$-input, $2^{n}$-output lookup table with learned 1-bit entries. Each permutation of the contents of this table represents a unique Boolean function, meaning there are $2^{2^{n}}$ possible functions for a single RAM node, many of which are nonlinear. By contrast, a single \texttt{XNOR}-and-popcount neuron in a BNN can only learn one of the $n*2^{n}$ linear functions of its inputs.

The downside of this expressiveness is that the size of a RAM node grows exponentially with its number of inputs, quickly becoming intractable. Therefore, layers in a WNN are typically composed of many RAM nodes which are each only sensitive to a subset of the layer's inputs.

Training a WNN entails learning Boolean functions in its component RAM nodes. Many approaches have been explored for this, including both supervised~\cite{10.5555/284803.284804} and unsupervised~\cite{unsupervised_wnn} techniques. Most typically, WNNs are trained using a supervised one-shot approach. All RAM nodes begin filled with zeros. Encoded inputs are presented sequentially to the network. When a RAM node receives a training input, it sets the corresponding location in its memory to 1. Note that presenting the same pattern to a node again has no further effect; thus, there is no advantage to having multiple epochs of training. By leveraging one-shot training techniques, WNNs can be trained up to four orders of magnitude faster than DNNs and other well-known computational intelligence models such as SVM~\cite{cluswisard}.

Since WNNs only learn patterns they were exposed to during training, one might expect them to have difficulty generalizing to new data. However, although an inference sample may not be identical to any training sample, many of the ``subpatterns'' seen by individual RAM nodes will have also been present in training data. Therefore, as long as an inference sample is not too different from any training sample, the network can still effectively generalize.

We have only described the fundamentals of WNNs here; many variants have been explored, and we refer the curious reader to~\cite{weightless_review} for a more in-depth discussion. We will however present one WNN model architecture, WiSARD, in detail, since it serves as the baseline for the improvements discussed in this paper.

Though not the focus of this paper, computer architects may note some similarities between WNNs and the predictors used in modern microprocessors. For instance, RAM nodes are conceptually similar to branch predictor tables, and using a concatenated input vector to index into a RAM node is conceptually similar to using a branch history register in a table-based branch predictor. Table updates in branch predictors can be viewed as an online version of the one-shot training rule.

%\subsection{WiSARD}

\begin{figure}[htbp]
\centerline{\includegraphics[width = 1.1\columnwidth]{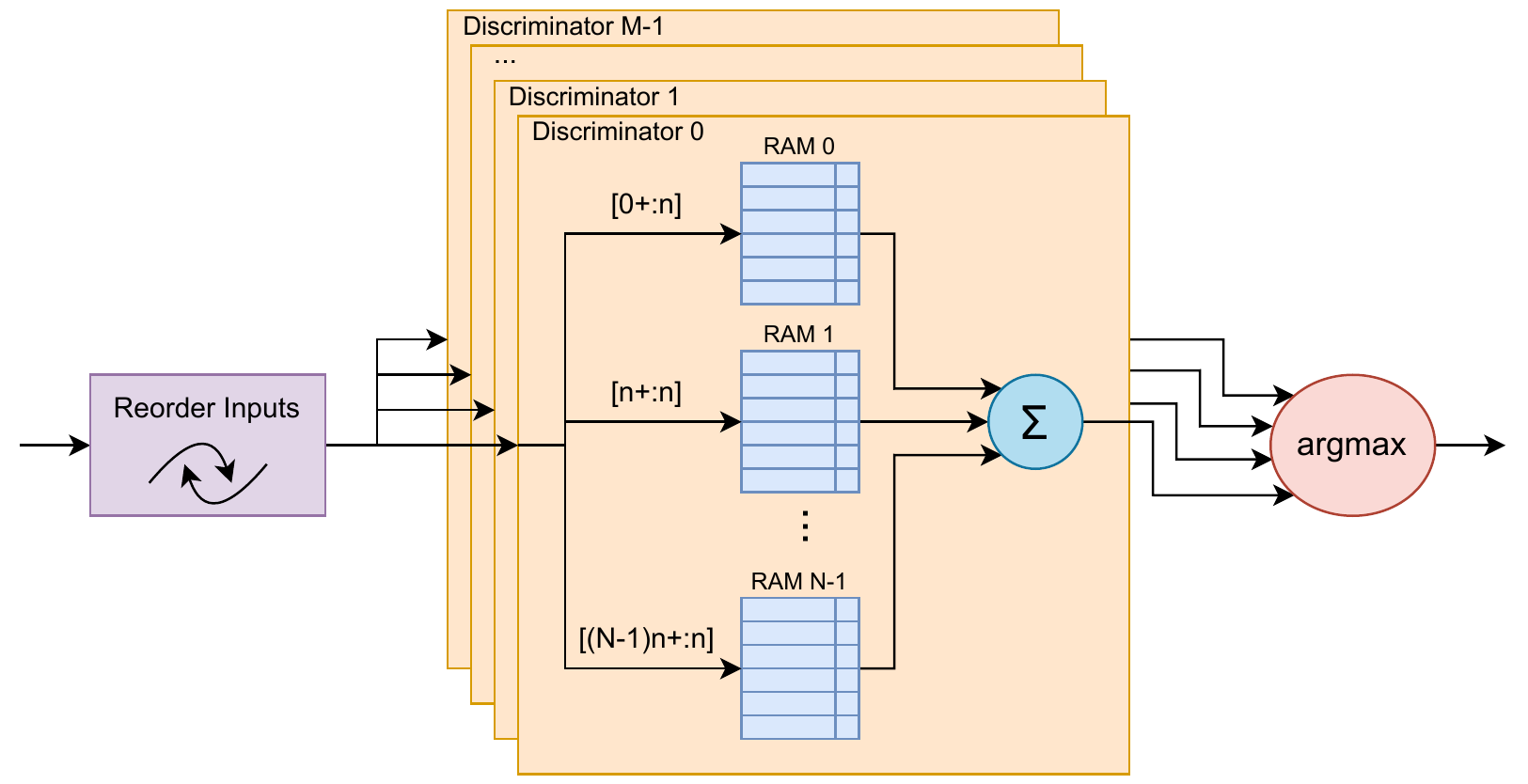}}
\caption{A depiction of the WiSARD WNN model with $M$ classes, $n$ inputs per RAM node, and $N$ RAM nodes per discriminator. This model has a total of $MN$ RAM nodes and $MN2^{n}$ bits of state.}
\label{fig:wisard}
\end{figure}

{\bf WiSARD:} While foundational research in WNNs began in the 1950s, the first WNN architecture to achieve broad success was WiSARD (Wilkie, Stonham, and Aleksander’s Recognition Device)~\cite{wisard}, introduced in 1981 and produced commercially from 1984. WiSARD was designed for classification tasks for purposes such as anomaly detection. A WiSARD model, as depicted in Figure \ref{fig:wisard}, is composed of submodels, known as \textit{discriminators}, which are each specialized for one of the possible output classes. These discriminators are in turn composed of $n$-input RAM nodes; for an $I$-input model, there are $N \equiv I/n$ such nodes per discriminator. Inputs to the network are assigned to RAM nodes using a pseudo-random mapping (also referred to as reordering); typically this same mapping is shared between all discriminators, meaning RAM nodes at the same index in different discriminators will have the same inputs.

\begin{figure}[t]
\centerline{\includegraphics[width = 0.95\columnwidth]{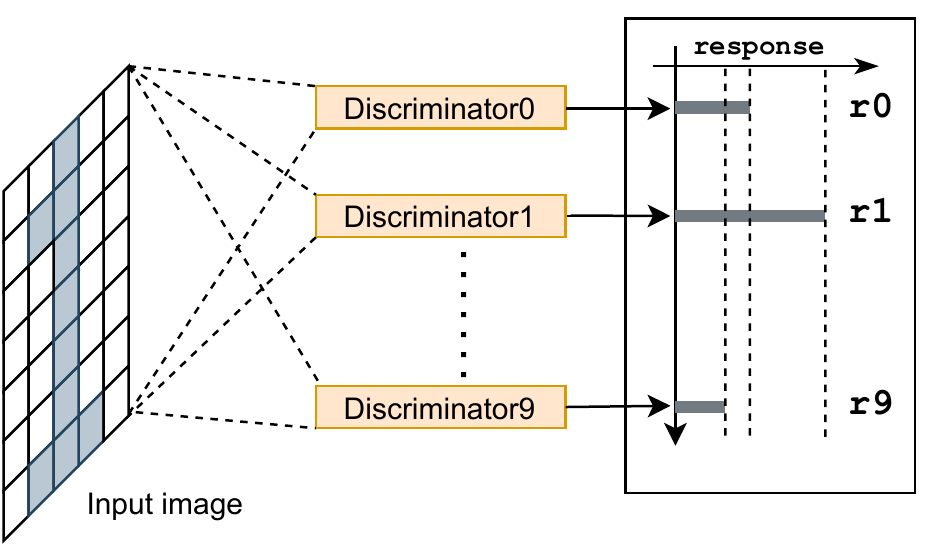}}
\caption{A WiSARD WNN model recognizing digits. In this example, the input image contains ``1'', and the corresponding discriminator produces the strongest response.}
\label{fig:wisard_basics}
\end{figure}

During training, inputs are presented only to the discriminator corresponding to the correct output class. During inference, inputs are first presented to all discriminators. The outputs of the RAM nodes in each discriminator are then summed to produce response values, and the class corresponding to the discriminator with the strongest response is taken to be the prediction of the network. Figure \ref{fig:wisard_basics} shows a simplified WiSARD model performing inference. In this example, the response from Discriminator 1 is the strongest since the input image contains the digit ``1''.

If an input seen during inference is identical to one seen during training, all RAM nodes of the corresponding discriminator will output 1, yielding the maximum possible response. On the other hand, if a pattern is similar but not identical to training patterns, then some subset of the RAM nodes may produce a 0, but many will still output 1. As long as the response of the correct discriminator is still stronger than the responses of any other discriminator, the network will output a correct prediction. In practice, WiSARD has a far greater ability to generalize than simpler WNN models.

WiSARD's performance is directly related to the choice of $n$. Small values of $n$ force the model to only learn simple patterns, which may improve its ability to generalize but can also prevent it from effectively learning. Large values of $n$ produce more specialized behavior, but may also result in overfitting to training data~\cite{wisard}. Recent results have formally demonstrated that the VC dimension\footnote{The Vapnik–Chervonenkis (VC) dimension measures the complexity of the knowledge represented by a set of functions that can be encoded by a binary classification algorithm~\cite{vc_dimension}. While usually approximated by statistical methods, it is possible to establish the exact VC dimension for some learning methods, including WiSARD.} of WiSARD is very large~\cite{wisard_vc}, indicating it has a large theoretical capacity to learn patterns.

\section{Proposed Design: \wname}
\label{sec:proposal}

\begin{figure}[t]
\centerline{\includegraphics[width = 0.5\textwidth]{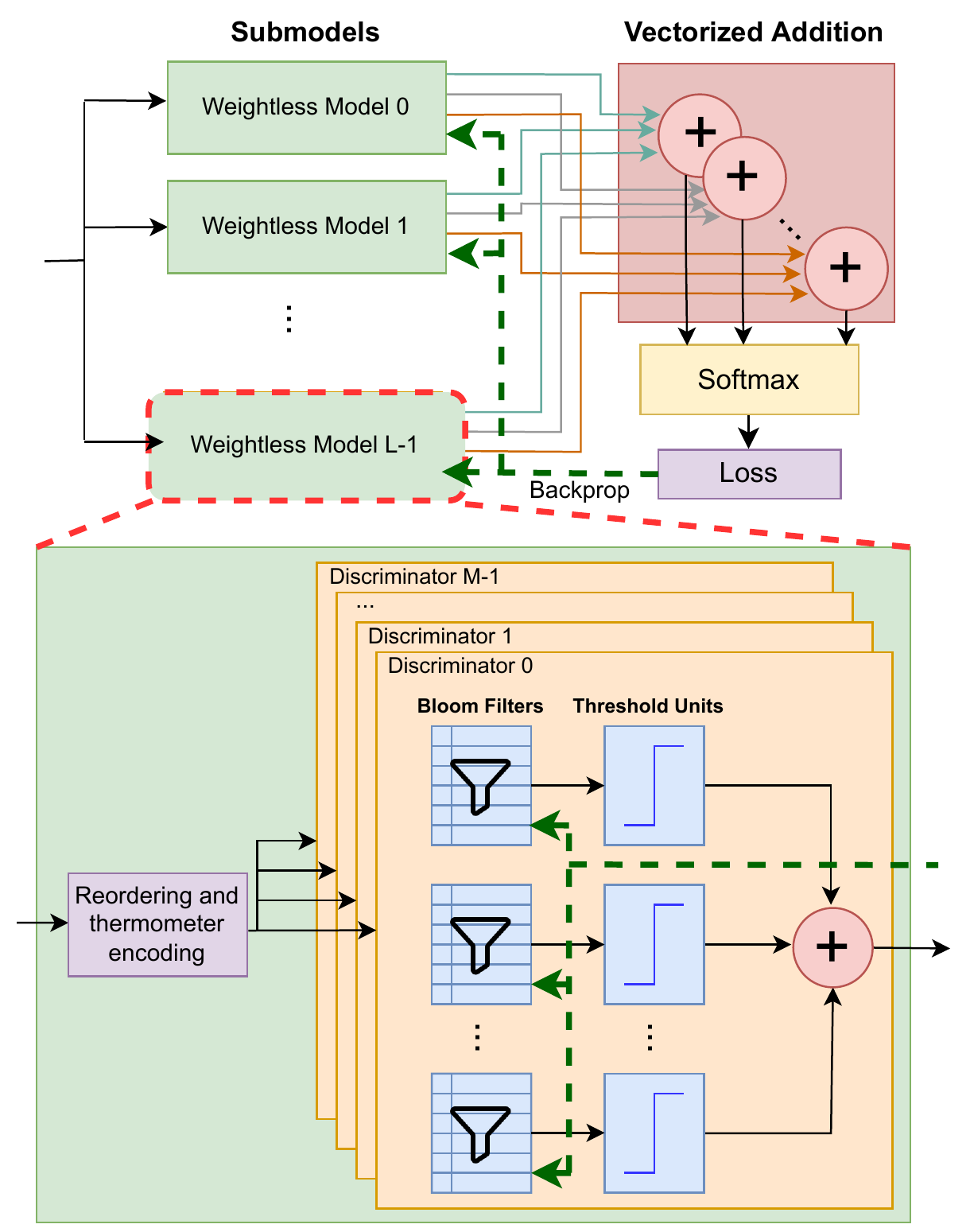}}
\caption{The \wname~model is composed of an ensemble of submodels, each of which is a WNN. Each submodel is composed of discriminators. Discriminators use continuous Bloom filters during training to allow for gradient-based weight updates. This figure shows the multi-shot training process, which uses backpropagation based on the straight-through estimator.}
%\caption{High-level overview of the \wname~model, demonstrating multi-shot training. The model is composed of an ensemble of submodels, which are themselves composed of discriminators. Discriminators use continuous Bloom filters during training to allow for gradient-based weight updates. Multi-shot training is a modified form of backpropagation based on the straight-through estimator.}
\label{fig:model_overview}
\end{figure}

%In this paper, we present \wname, a WNN architecture \revision{which improves on the prior work by incorporating} several enhancements: (i) backpropagation-based multi-shot training (ii) ensemble models (iii) counting Bloom filters (iv) inexpensive arithmetic-free hash function to address RAM node discriminators, and (v) a Gaussian-based non-linear thermometer encoding to improve model accuracy.
%We also present an FPGA-based accelerator for this architecture, targeting low-power edge devices, shown in Figure \ref{fig:proposed_design}.
%We incorporate both hardware and software improvements over prior work.
The \wname~model includes several enhancements over the baseline WiSARD model: (i) counting/continuous Bloom filters, (ii) non-linear thermometer encoding, (iii) ensembles, and (iv) pruning. Furthermore, \wname~models can be trained in two ways: (i) an enhanced version of the conventional one-shot technique or (ii) a novel multi-shot training algorithm, inspired by BNNs and based around backpropagation and the straight-through estimator.

\subsection{Model Overview}

Figure \ref{fig:model_overview} shows the \wname~model at a high level, including training with the multi-shot technique. 
Inputs are fed to an ensemble of smaller models. We find that ensembles of small models give better accuracy than monolithic large models, and incorporate this into \wname's design. The aggregation of the results from individual models is done in the ``Vectorized Addition'' block in the figure. During multi-shot training, we take a softmax of the outputs of the aggregation and compute cross-entropy loss.

A multi-bit thermometer encoding is used to represent inputs with greater granularity. RAM nodes are implemented using hash-based Bloom filters for compression. During multi-shot training, Bloom filters internally hold floating-point values and are binarized using a unit step function.

The multi-shot training technique for \wname~is a modified version of backpropagation based on the straight-through estimator~\cite{straight_through_estimator}. The flow of gradients during backpropagation is shown with the green dotted arrow in the figure; the threshold units are treated as identity functions and ignored. Pruning, which eliminates RAM nodes which contribute least to overall accuracy, is a post-training technique and is therefore not shown in Figure \ref{fig:model_overview}.

\subsubsection{Counting/Continuous Bloom Filters} \label{sec:bloom_filter}
A major challenge impacting WiSARD is its exponential growth in model size as the number of inputs to each RAM node increases. However, in practice, these large RAM nodes learn Boolean functions with few minterms, meaning their contents are sparse. In Bloom WiSARD~\cite{Arajo2019MemoryEW}, it was demonstrated that hash-based data structures such as Bloom filters can be used to reduce the data footprint of RAM nodes. 
%Lizy commented line 22
Bloom filters consist of a lookup table and $k$ independent hash functions, and output 1 only if all $k$ hashed locations corresponding to an input are 1. Since no collision checking occurs, Bloom filters may produce false positives. However, this was found to have minimal impact on accuracy.

%A Bloom filter\cite{BloomFilter} is a hash-based data structure for approximate set membership. For a given set $S$ and value $x$, a Bloom filter may yield one of two responses: ``possibly $x \in S$'' and ``definitely not $x \in S$'' - in other words, they may have false positives, but not false negatives. Internally, a Bloom filter is composed of a lookup table and $k$ independent hash functions. To add a value $x$ to $S$, $x$ is hashed using all functions and the corresponding $k$ locations in the LUT are set. To test $x \in S$, the boolean \texttt{AND} of the $k$ corresponding locations is computed. Bloom filters were previously used instead of LUTs as RAM nodes in the aptly named Bloom WiSARD model~\cite{Arajo2019MemoryEW}, where they were found to greatly decrease model size with a minimal impact on accuracy. However, the Bloom filters used in this paper had notable limitations.
In order enable more sophisticated training approaches, we use two variants of the Bloom filter: \textit{counting} and \textit{continuous} Bloom filters. After training, the contents of these filters are binarized, and they are replaced with conventional Bloom filters for inference.
Figure \ref{fig:model_overview} shows the substitution of enhanced Bloom filters in place of conventional RAM nodes in the discriminators. 

\begin{figure}[t]
\centerline{\includegraphics[width = 0.9\columnwidth]{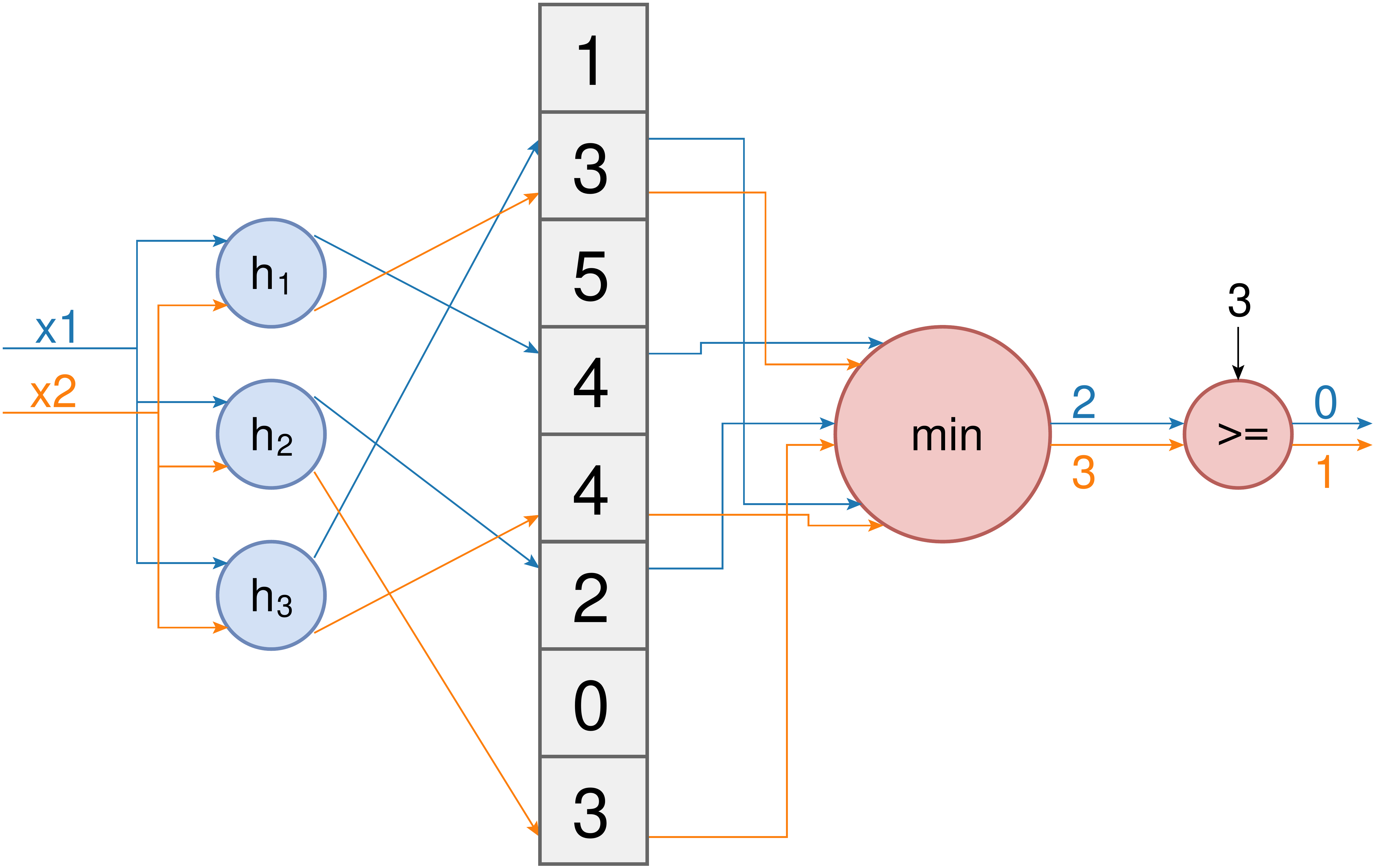}}
\caption{An example of a counting Bloom filter with three hash functions and threshold $b=3$.}
\label{fig:counting_bloom_filter}
\end{figure}

When training a WiSARD model using conventional Bloom filters, RAM entries are set the first time a pattern is seen during training, and are not subsequently updated. The issue with this approach is that it treats rare patterns as equally important to common ones. Bleaching~\cite{bleaching} is a technique in which patterns which were seen fewer than some threshold $b$ times during training are ignored during inference. Typically, this is achieved using tables of counters. In \wname, while training using the one-shot strategy, we replace Bloom filters with counting Bloom filters, which use multi-bit counters instead of single-bit entries, enabling bleaching and hashing in the same model. During training, whenever a pattern is seen, the smallest of its corresponding counter values is incremented (multiple counters in the event of a tie). Figure \ref{fig:counting_bloom_filter} shows the behavior of a counting Bloom filter during inference. The input to the filter is hashed, and these hashes are used to access counter values.
For input $x_{1}$, this corresponds to counter values $\{3,\,4,\,2\}$; for $x_{2}$, this corresponds to $\{3,\, 4,\, 3\}$.
The smallest accessed counter value is then found (2 for $x_{1}$; 3 for $x_{2}$) and compared against a threshold value $b$ to determine the response of the filter. For this example, the threshold $b=3$, so $x_{1}$ produces an output of 0 and $x_{2}$ an output of 1. Filter responses become ``possibly seen at least $b$ times'' and ``definitely not seen $b$ times''.

When training using the multi-shot strategy, we go further by replacing Bloom filter entries with floating point values. These \textit{continuous} Bloom filters produce an output of 1 if the smallest entry is at least 0. Using continuous Bloom filters allows us to train using gradient-based techniques.

Since the hash functions used for our Bloom filters do not need to be cryptographically secure and have constant-length input, we are able to use simple \textbf{arithmetic-free} hash functions. In particular, we use functions drawn randomly from the H3 family~\cite{CARTER1979143}, a set of hash functions which consist of simple logical operations and differ from each other only by the choice of a random parameter. Bloom WiSARD derived its independent hash functions using a double-hashing technique based on the MurmurHash~\cite{MurmurHash} algorithm. Though simpler than cryptographic hash functions, MurmurHash is designed to handle variable-length inputs and requires substantial arithmetic. 
%\textbf{arithmetic free hashing} % Incorporate this!

%\subsubsection{Backpropagation-based Training}

%Traditionally WNNs use one-shot training because presenting the same input to a node again has no effect usually, since the corresponding bit position has already been set. However, a one-shot training has severe accuracy limitations. We develop a methodology to use gradient descent algorithms to train the RAM node entries. During training we use a floating point number in the range -1 to +1. A step function 

\subsubsection{Gaussian Nonlinear Thermometer Encoding}

Inputs to WNNs are traditionally represented as 1-bit values: 1 if an input is greater than its mean in the training data and 0 otherwise. However, the granularity that can be provided using a single bit is very limited. Integer encodings are not a good choice, since different bits carry different amounts of information, meaning the least significant bits are essentially noise when used to form an address. Thermometer encodings are the preferred multi-bit encoding in prior work~\cite{wisard_encoding}.

\begin{figure}[hbtp]
\centerline{\includegraphics[width = 0.55\columnwidth]{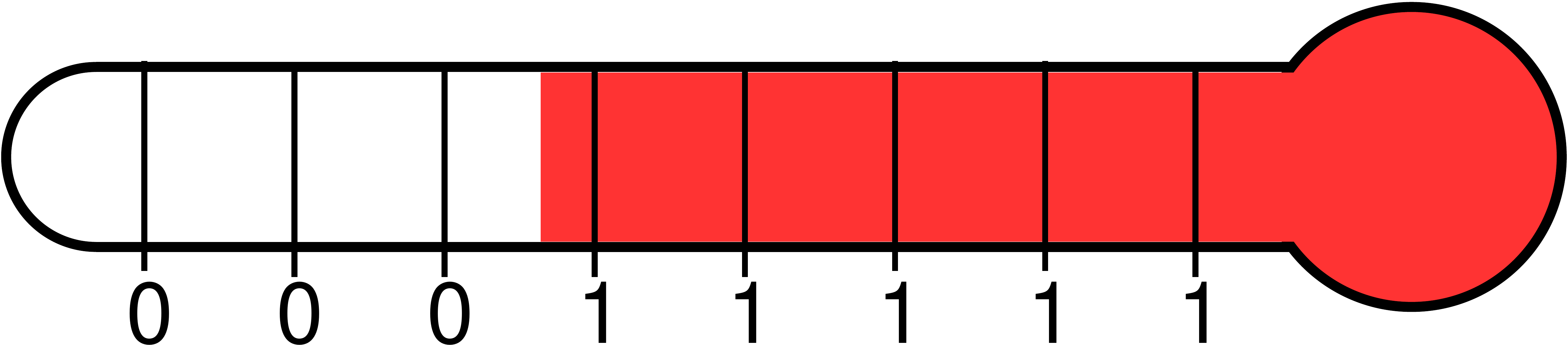}}
\caption{Like the mercury passing the gradations in a thermometer, in a thermometer encoding, bits are set to 1 from least to most significant as the encoded value increases.}
\label{fig:thermometer}
\end{figure}

Thermometer encoding is a \textit{unary} coding in which a value is compared against a series of increasing thresholds, with the $i$'th bit of the result representing the comparison against the $i$'th threshold. As Figure \ref{fig:thermometer} shows, input values behave like the mercury in an analog thermometer: as a value increases and passes thresholds, bits are set from least to most significant.

Most prior work using thermometer encodings uses equal intervals between the thresholds. The disadvantage of this approach is that a large number of bits may be dedicated to encoding outlying values. In \wname, we instead assume that each input follows a normal distribution, and compute its mean and standard deviation from training data. For a $t$-bit encoding, we use thresholds to divide the Gaussian into $t+1$ regions of equal probability. This Gaussian encoding provides increased resolution for values near the center of their range. Gaussian thermometer encodings were also explored in~\cite{Xavier2020DetectionOE} in a narrow scope; we use them more generally, showing they are useful even when the underlying data is not actually Gaussian.

\subsubsection{Efficient Ensembles}
\label{sec:ensemble}

\begin{figure}[t]
\vspace{-10pt}
\centerline{\includegraphics[width = 0.9\columnwidth]{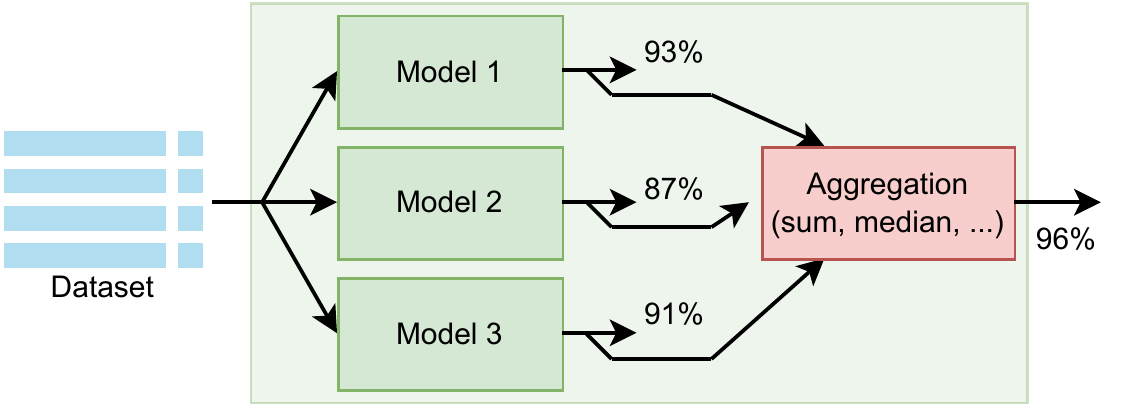}}
\caption{Simplified view of an ensemble model. An ensemble combines multiple weak classifiers to create a single stronger classifier.}
\label{fig:ensemble}
\end{figure}

Ensembles, as shown in Figure \ref{fig:ensemble}, combine multiple weak classifiers into a single strong classifier. Ensembles have been extensively studied in machine learning, and are the driving concept behind machine learning techniques such as Bayesian averaging, Boosting, and Bagging~\cite{ensemble_review}.

One consideration when designing an ensemble model is how the predictions of multiple submodels should be combined into a single output. As Figure \ref{fig:model_overview} shows, \wname~composes submodels by summing the response values for each discriminator across the submodels. In other words, a model with $L$ submodels and $M$ classes will produce an $L\times M$ matrix of responses $R$, and a top-level $M$-element response vector $\vec{r} \triangleq R^{\intercal} \vec{1}$.

One might reasonably expect that using ensembles would increase total model size, since the number of RAM nodes is much greater. In practice, we have found this to not necessarily be the case - recall that the key strength of ensembles is their ability to combine weak classifiers. This means that individual submodels of an ensemble can be made much smaller than standalone models without harming ensemble accuracy. 

In an ensemble of submodels with different numbers of inputs per filter, filters with many inputs may also selectively ``ignore'' simple patterns where most inputs are don't-cares, relying on filters with fewer inputs to capture these patterns. This helps reduce the number of table entries needed for Bloom filters.

Since the amount of hashing required for inference increases linearly with the number of submodels, we avoid using ensembles with many submodels. We also do not train ensembles with the one-shot training rule, since this rule gives poor accuracy when small Bloom filters have many inputs.

Ensembles of WiSARD models were investigated in~\cite{ensemble_wisard}, but without backpropagation or multi-shot training, and resulted in large models with marginal accuracy benefit.
%prior work increased complexity by orders of magnitude for a marginal reduction in error. \wname~is the first WNN to use ensembles efficiently.
%Prior work~\cite{ensemble_wisard} experimented with using the weighted arithmetic mean, harmonic mean, and median for aggregation. Our approach is functionally equivalent to an \textit{unweighted} arithmetic mean (since the sum is the mean times a constant scaling factor), and does not require multipliers, dividers, or sorting networks, unlike the prior WNN ensemble aggregation functions.
%used an alternative aggregation method based on ``confidence scores'': the difference between the highest and second-highest responses of a submodel. In this approach, the submodel with the highest confidence was chosen as the predictor. We experimented with this technique as well, but found it to be both computationally expensive and less accurate.

\subsubsection{Pruning} \label{sec:pruning}
Pruning removes parameters and connections from a model to reduce its size and complexity. Optimized DNNs~\cite{deep_compression,scalpel,dnn_weight_pruning_framework,patdnn,once_for_all} have applied pruning techniques with excellent results, but similar concepts have not been used in prior WNNs.

After training, the correlations between RAM node outputs and correct model outputs are calculated for each RAM node in the model. A fixed fraction of the RAM nodes with the lowest correlation in each discriminator are removed from the model. Since this pruning reduces the maximum possible response of a discriminator, we next learn a set of integer biases which are added to the outputs of the discriminators. Finally, we fine-tune the remaining RAM nodes using the multi-shot training process.

In our experiments, we observed that we could typically prune around 30\% of RAM nodes, reducing model size proportionately, with a minimal impact on accuracy. Though this does not approach the degree of pruning which is frequently possible for DNNs, it is still a significant reduction in \wname's memory requirement. When used in an ensemble, the bias can be summed across the submodels, meaning the only overhead for pruning is a single addition.

\subsection{Training \wname}
We use both one-shot and multi-shot training techniques for \wname. The one-shot technique is similar to how WiSARD models are traditionally trained. The multi-shot technique provides better accuracy, particularly for larger datasets, and is necessary in order to use ensemble models and pruning, but does require much more training time than the one-shot approach.

\subsubsection{One-shot Training}
\begin{figure}[t]
\vspace{-20pt}
\centerline{\includegraphics[width = 1.05\columnwidth]{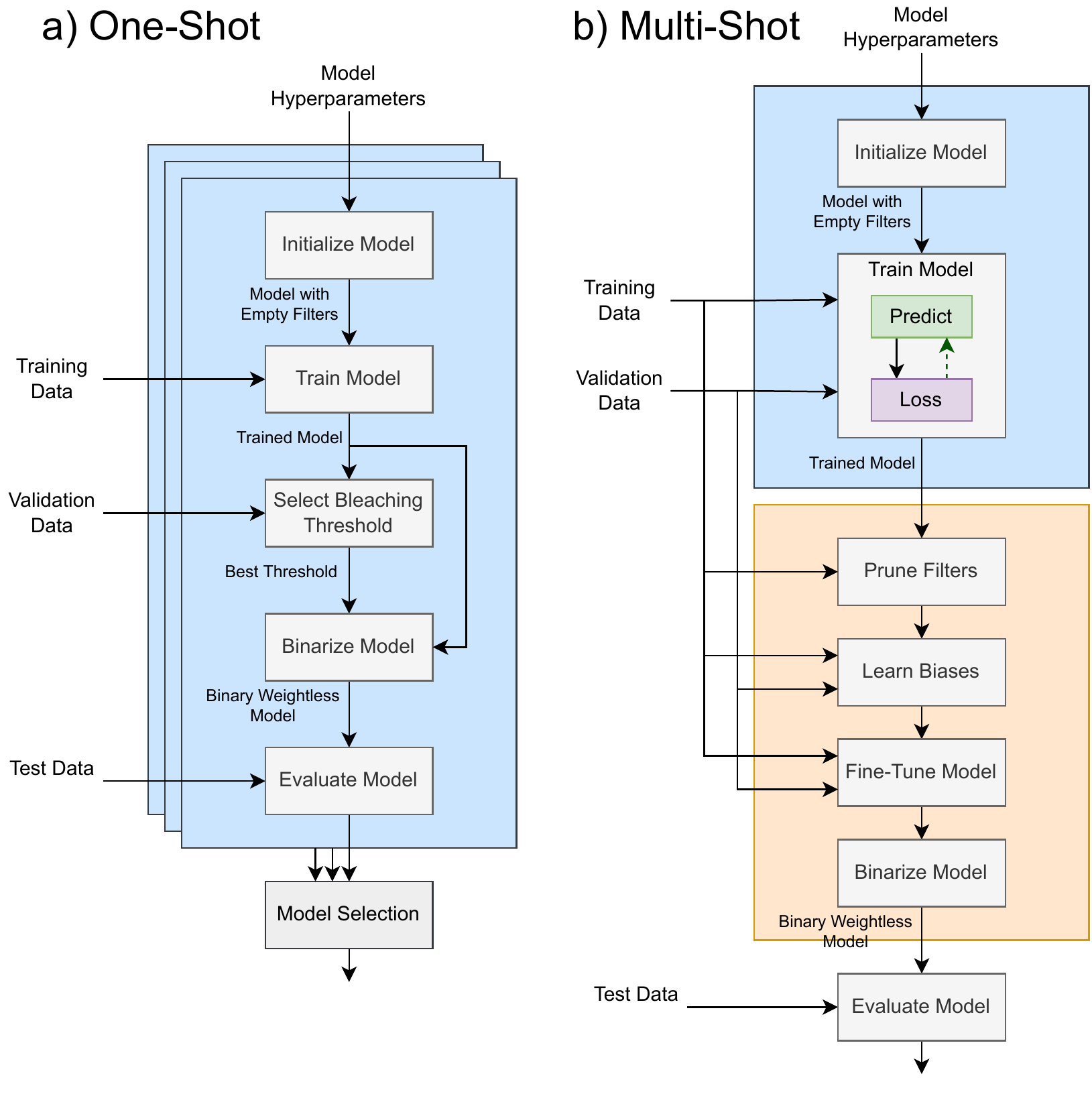}}
\vspace{-10pt}
\caption{\wname~supports both one- and multi-shot training techniques. (a) In one-shot training, encoded training samples are sequentially presented to the model and used to update counter values. Afterwards, a bleaching threshold is optimized to eliminate rare patterns. (b) In multi-shot training, a gradient-based update rule is used to learn counter values over multiple epochs. Next, weakly correlated filters are pruned and replaced with a constant bias, and the remaining filters are fine-tuned.}
\label{fig:training}
\vspace{-15pt}
\end{figure}

The one-shot training technique is a computationally simple approach which is similar to how prior WNNs were trained. The process is summarized in Figure \ref{fig:training}a. Hyperparameters include the thermometer encoding and the configuration information for the Bloom filters (number of inputs, table size, and number of hash functions). During training, encoded input samples are sequentially presented to the discriminator corresponding to the correct output class. Counting Bloom filters update their contents by incrementing counters according to the method discussed previously (Section \ref{sec:bloom_filter}).

After training, patterns that were rarely seen during training are discarded using bleaching~\cite{bleaching}. Bleaching finds a value $b$ such that all patterns seen fewer than $b$ times during training are discarded. If bleaching is not used, almost all RAM node entries may become 1 given a large training dataset, harming generalization. This behavior is known as \textit{saturation}. WNNs which do not use bleaching need careful manual feature extraction and data selection to avoid saturation.

We use a binary search strategy to find the value of $b$ which maximizes model accuracy on the validation set. Afterwards, all filter values less than $b$ are replaced with 0, and all remaining filter values are replaced with 1. This allows the counting Bloom filters to be replaced with conventional binary Bloom filters for inference, simplifying the hardware.

\subsubsection{Multi-shot Training}
While the one-shot technique is efficient, it is ultimately limited by its inability to consider feedback during training. However, conventional gradient-based techniques do not work for WNNs since the outputs of Bloom filters are binary rather than continuous. A similar problem also impacts BNNs, which led to the development of gradient-based training algorithms that worked with binary weights. We base our multi-shot training technique on the BNN training algorithm described in~\cite{binary_net}.

\wname~uses continuous Bloom filters, with floating-point entries between -1.0 and 1.0, for training with the multi-shot technique. During the forward training pass, entries are binarized using a unit step function:
\[
    f(x) = \begin{cases}
        0 & x < 0\\
        1 & x \ge 0
    \end{cases}
\]
Backpropagation does not work with the unit step function, since its derivative is 0 everywhere except $x=0$ (where it is infinite). However, when computing gradients, we instead treat the unit step function as the identity function $f(x) = x$, meaning $f^{\prime}(x) = 1$. This technique, known as the \textit{straight-through estimator}, has been used for training low-precision and binary networks~\cite{straight_through_estimator}, but we are the first to use it for WNNs. Since our model is only a single layer, we do not have to propagate gradients through the indexing and hash operations.

Figure \ref{fig:training}b summarizes our multi-shot training process. After training, models are pruned and fine-tuned according to the method discussed in Section \ref{sec:pruning}. After pruning, the continuous Bloom filters are binarized by applying the unit step function.

Models were trained using the Adam optimizer~\cite{adam} with base learning rate $10^{-3}$. Weights were initialized with uniform distribution $\mathcal{U}(-1,\,1)$. To prevent overfitting, we added dropout regularization~\cite{dropout} ($p=0.5$) to the outputs of the filters. For the MNIST\cite{lecun-mnisthandwrittendigit-2010} dataset, we also experimented with a simple form of data augmentation: we made 9 copies of each image in the training set, shifted horizontally and vertically between -1 and 1 pixels.

\subsection{Inference Accelerator Architecture}
\begin{figure*}[t]
\centerline{\includegraphics[width = 0.9\textwidth]{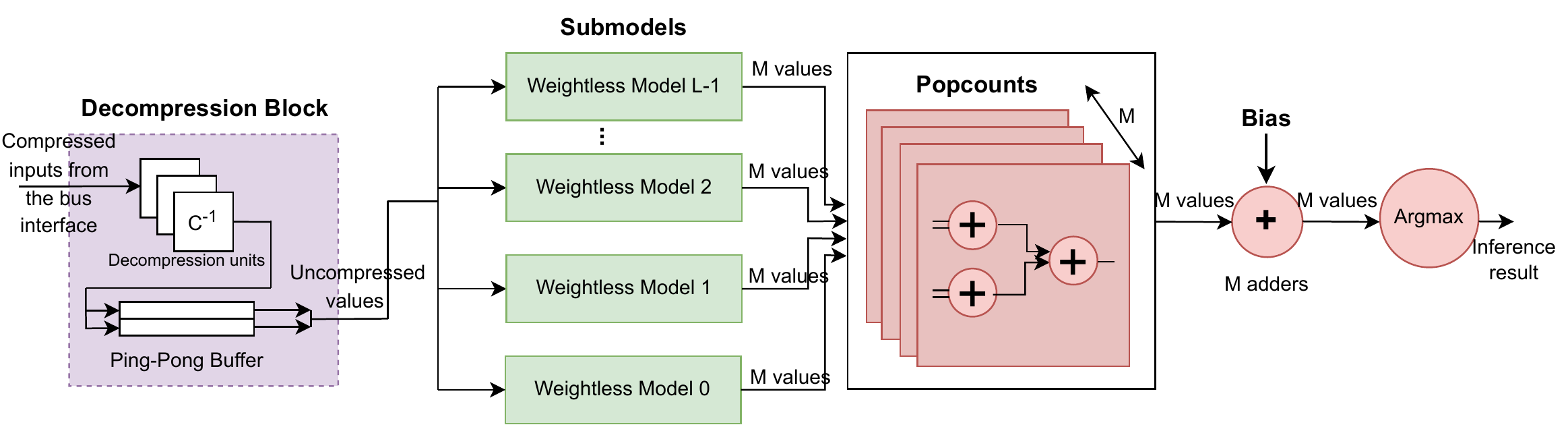}}
\caption{A diagram of the \wname~inference accelerator architecture. Input is deserialized and, if needed, decompressed, before being passed to an ensemble of submodels. The outputs of the submodels are summed and biased to get per-class responses, and the index of the strongest response is taken as the prediction.}
\label{fig:proposed_design_a}
\end{figure*}

\begin{figure*}[t]
\centerline{\includegraphics[width = 0.9\textwidth]{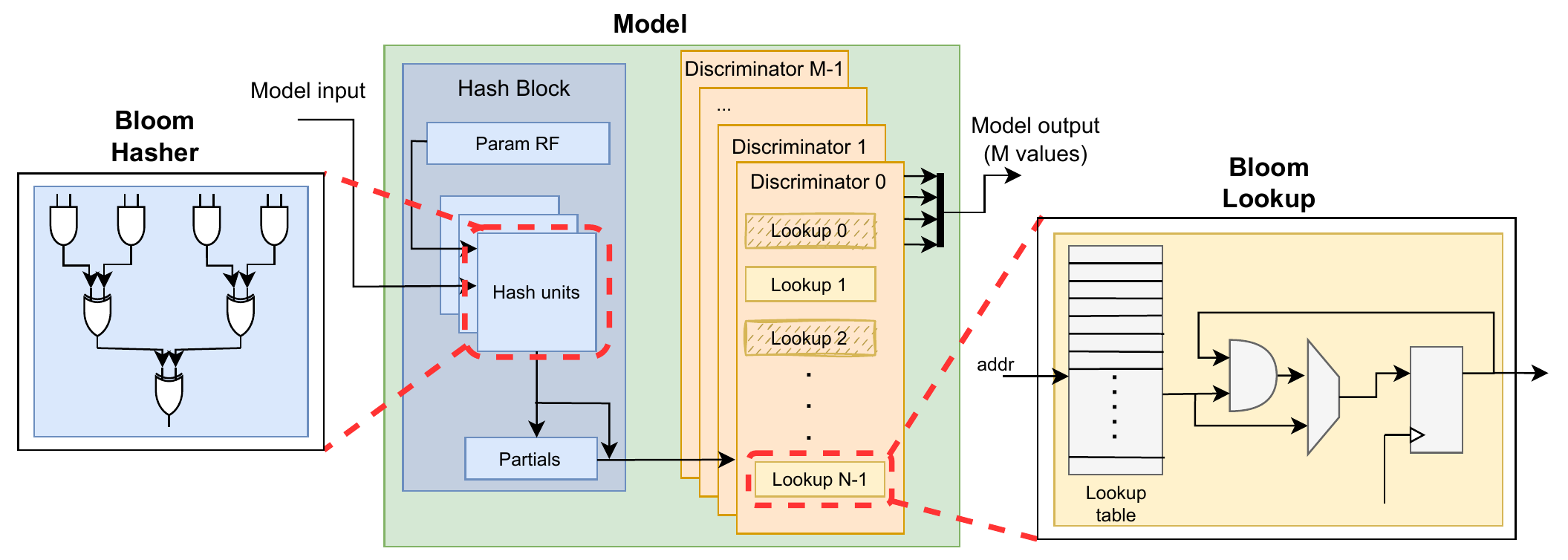}}
\caption{Details of each submodel in the \wname~inference accelerator. Each submodel contains a hardware block for computing hash functions, and a number of hardware units for performing lookups. These blocks collectively compose the Bloom filters, which are divided into separate units to eliminate redundant computation. }
\label{fig:proposed_design_b}
\end{figure*}

Figures \ref{fig:proposed_design_a} and \ref{fig:proposed_design_b} show the block diagram of our pipelined \wname~inference accelerator. 
%Like most accelerators targeting the very low-power edge space, we do not provide support for on-chip training.
In order to simplify control logic, units on the chip operate in lockstep to the greatest extent possible. This means that an entire input sample must be read in before computation can begin. This deserialization is performed by the bus interface (not shown in the figure). Input data may optionally be compressed by replacing unary thermometer-encoded values with binary values representing how many bits are set. This reduces data movement from off-chip, but requires a decompression unit to recover the thermometer encoding, shown on the left in Figure \ref{fig:proposed_design_a}. This unit is eliminated from the design if input data is not compressed.

The discriminators in Figure \ref{fig:proposed_design_b} exhibit several optimizations. Though each H3 hash function in a Bloom filter requires a different random parameter, there is no disadvantage to sharing these parameters between all Bloom filters in a submodel. Therefore, a central register file (shown as ``Param RF'' in the figure) is used to store hash parameters. Since input order is shared between discriminators in a submodel, when hash parameters are also shared, all discriminators will receive the same hashed values. Therefore, it is redundant to calculate hashes for each discriminator separately; instead, we use a single central hashing block. The hash block is itself composed of many pipelined hash units which process input sequences with a throughput of 1 hash/cycle. As shown in the left part of Figure \ref{fig:proposed_design_b}, these hash units perform only \texttt{AND} and \texttt{OR} operations, with no arithmetic component. If hashing is fully parallelized, the performance of the design will be heavily bottlenecked by off-chip bandwidth. Therefore, we reduce the number of hash units to the minimum number sufficient to achieve maximum throughput, and accumulate partial hash results in a buffer. Once the buffer is full and the last partial hash is available, all Bloom filters perform a lookup in lockstep.

Since hashing is moved into a central block, discriminators themselves contain only lookup units. The lookup unit, shown at the right of Figure \ref{fig:proposed_design_b}, consists of a lookup table and the hardware to perform an \texttt{AND} reduction. A 1-bit accumulator in each lookup unit can take as input either the output of the LUT or the \texttt{AND} of that output and its current contents. Once all hash lookups have been performed, the outputs of the lookup units are marked as valid.

Each submodel in an ensemble must compute its own hashes, since input orders and hash input and output widths vary. Since different submodels have different table contents, sizes, and pruning, they also have their own sets of filter units.

Popcounts and submodel response accumulation are performed using a vector of adder trees, shown in the right of Figure \ref{fig:proposed_design_a}. Bias values are added to these results (Section \ref{sec:pruning}), and the index of the strongest response is computed to produce a final inference result.

\section{Evaluation Methodology}
\label{sec:methodology}

%\subsection{Benchmarks}

%We use the MNIST dataset for our evaluation.

We compare \wname~against state-of-the-art BNNs, quantized DNNs, and WNNs, using designs in FPGAs, ASICs and software models. The methodology for the various comparisons is described in this section. 

The FINN~\cite{finn_documentation, finn} framework from Xilinx Research Labs for BNN inference on FPGAs is used as a comparison for our FPGA implementation. We use the SFC-max, MFC-max and LFC-max networks described in~\cite{finn}. 
These are 3-layer fully-connected network topologies for classifying the MNIST dataset~\cite{lecun-mnisthandwrittendigit-2010}, with different numbers of neurons (SFC contains 256 neurons per layer, MFC contains 512 neurons per layer and LFC contains 1024 neurons per layer). The ``max'' suffix indicates that these are performance-optimized designs intended to achieve high peak throughput.
To the best of our knowledge, these are most efficient implementations of BNNs on FPGAs for MNIST.

We compare our ASIC implementation against Bit Fusion~\cite{bit_fusion_documentation, bitfusion:isca:2018}, a quantized DNN hardware architecture which supports dynamic precision. 
%Bit Fusion can accelerate quantized DNNs efficiently. 
A highly quantized LeNet-5 architecture (2-bit; Ternary) is used by Bit Fusion for classifying MNIST. This NN architecture is proposed in~\cite{TWN} and shown to have accuracy close to the original LeNet5. 
We use a Bit Fusion design with precision optimized for this model; i.e. minimum and maximum precision are 2 bits. We identify three efficient variations of the accelerator for the workload, with configuration parameters:
\begin{itemize}[leftmargin=*]
 \setlength\itemsep{0pt}
 \item[] \{S=8x8, WBUF=32KB, ABUF=16KB, OBUF=8KB\}
 \item[] \{S=16x16, WBUF=64KB, ABUF=32KB OBUF=16KB\}
 \item[] \{S=32x32, WBUF=64KB, ABUF=32KB, OBUF=16KB\}
\end{itemize}
%We consider 4 different Bit Fusion accelerator variations for comparisons -
%\{S=16x32,Pmax=8,Pmin=2\}, \{S=16x32,Pmax=2,Pmin=2\}, \{S=16x16,Pmax=8,Pmin=2\}, %\{S=16x16,Pmax=8,Pmin=2\}. 
Here, `S' denotes the size of the systolic array in the Bit Fusion core, and WBUF, ABUF, OBUF denote the size of the weight, activation and output buffers respectively.
%and Pmax and Pmin denote the maximum and minimum precisions supported on the accelerator. 
%These variations are refered to as BF32-8, BF32-2, BF16-8 and BF16-8 respectively in Section \ref{sec:results}.
We refer to these variations as BF8, BF16 and BF32 respectively in Section \ref{sec:results}.
To implement the Bit Fusion accelerator, we obtained the RTL from the authors~\cite{bitfusion:isca:2018} and synthesized it using Cadence RTL Compiler 2017 and the FreePDK45~\cite{freepdk45} library. The Bit Fusion simulator 
~\cite{bit_fusion_documentation}, which combines the energy for the core reported by Cadence with the energy for RAMs/buffers reported by Cacti~\cite{cacti}, was used to evaluate the total energy spent on inference.

In order to gauge the impact of our improvements versus prior WNNs, we compared \wname~model sizes and accuracies against 
Bloom WiSARD~\cite{Arajo2019MemoryEW} on the same nine multi-class classification datasets originally used to evaluate that model. For datasets which did not have explicit train and test sets, we performed a 2:1 train/test split.
%This prior state-of-the-art WNN uses Bloom filters however {\color{red}not bleaching techniques.}
%to achieve far smaller model sizes than conventional WiSARD models with only slight penalties to accuracy.
%\cite{Arajo2019MemoryEW} was used for comparison against prior WNNs. 
%In \cite{wisard_fpga}, a WNN accelerator for MNIST was implemented on the Zybo FPGA (xc7z020clg400-1) with Vivado HLS. We perform a model-level comparison against both these WNNs.
%{\color{red}Do we want to say anything like XX was obtained...}

\subsection{Software Implementation}
%{\color{red} We first explored the hyperparameter space for submodels using the traditional one-shot training of WNNs.}
We implemented one-shot training in Python, using the Numba JIT compiler to optimize performance-critical sections. This code is CPU-only and single-threaded, but we used a machine with 64 cores (2x Intel E5-2698 v3) to train many models in parallel. We used a semi-automated sweeping methodology to identify optimal model hyperparameters. In particular, we swept over ranges of values for inputs, entries, and hash functions per Bloom filter and thermometer encoding bits per input. The results of this sweep were then used to identify models which achieved a good balance between accuracy and size.

%{\color{red}Ensemble model identification and the multi-shot training flow required for \wname} 
The multi-shot training flow is considerably more computationally intensive, as it requires gradient computation in addition to multiple epochs of training. However, it provides increased accuracy, and is necessary to use ensembles and pruning. We implemented multi-shot training using the PyTorch machine learning framework, and ran training on an NVIDIA Tesla M40 GPU. Forward and backward passes for Bloom filters were implemented as a single multi-dimensional gather/scatter operation, enabling efficient memory-parallel computation. Despite these optimizations, training times were much longer than with the one-shot algorithm. We believe this algorithm could benefit greatly from the much higher memory bandwidth and larger caches available on newer GPUs.

\begin{comment}
\subsection{Datasets and Training}
We created one-shot models for all classifier datasets discussed in~\cite{Arajo2019MemoryEW}: MNIST~\cite{lecun-mnisthandwrittendigit-2010}, Ecoli \cite{ecoli_dataset}, Iris\cite{iris_dataset},  Letter\cite{letter_dataset}, Satimage\cite{satimage_dataset}, Shuttle\cite{shuttle_dataset},  Vehicle\cite{vehicle_dataset}, Vowel\cite{vowel_dataset}, and Wine\cite{wine_dataset}. We performed a 90-10 split on the datasets, using the former portion to set the values in the counting Bloom filters and the latter portion to learn the optimal bleaching threshold.

While we experimented with using the multi-shot approach for several of these datasets, we ultimately found that it provided little to no benefit for small datasets. We speculate that, in the one-shot model, bleaching acts as a regularization technique. In the multi-shot model, which does not use bleaching, the model is able to overfit to artifacts in the training set with small datasets. Therefore, we only discuss results for the MNIST dataset with this technique.
\end{comment}

\subsection{Hardware Implementation}
The \wname~accelerator source is written using Mako-templated SystemVerilog. Mako~\cite{mako} is a template library which allows the Python interpreter to be run as a preprocessing step for code in arbitrary languages. We used Mako to automatically extract parameters from trained models, construct state machines, and decide how many functional units to instantiate in order to hit throughput targets while minimizing area. This templating allowed us to generate new accelerators just by changing command-line parameters. We simulated our designs using Synopsys VCS 2018 to ensure functional correctness and to evaluate the latencies and throughputs of our accelerators.

%We simulated both FPGA and ASIC deployments of our design. In particular, we targeted the FPGA of the Xilinx Zynq-7000 SoC ZC706 and used the FreePDK45 library for our ASIC simulations.
%We targeted two different Xilinx FPGAs for this project. For most designs, we used the xc7z020clg400-1 FPGA, a small, inexpensive FPGA available in the Zybo Z7 development board, which was used for prior work \cite{wisard_fpga}. For our largest design, we targeted the Kintex UltraScale xcku035-ffva1156-1-c. Timing, power, and area numbers were obtained from Xilinx Vivado. %Unlike that implmementation, however, we are only interested in Vivado-supplied timing, power, and area numbers, rather than whole-system performance. 

For our FPGA implementation and comparison, we use the Zynq Z7045 SoC platform, which was also used for FINN. Our design has the same I/O interface width as FINN (112 bits).
We used Xilinx Vivado 2019.2 to synthesize and implement our designs. We targeted the same frequency of operation (200 MHz) as FINN, though we were unable to achieve this frequency in all cases due to FPGA routing congestion. Resource usage and power consumption were obtained from Vivado reports.

For the ASIC implementation and comparison, we use the same tool and library as Bit Fusion for a fair comparison (FreePDK45~\cite{freepdk45} and Cadence RTL Compiler 2017). Our designs also have the same 192-bit interface width. The frequency of operation is 500 MHz, the same as the Bit Fusion designs. We used area and power numbers reported by Cadence RTL Compiler (there are no RAM blocks in our design, so using Cacti is not required, unlike with Bit Fusion).

\section{Results}
\label{sec:results}

\subsection{Selected \wname~Models}
We identified three \wname~models (ULN-S, ULN-M, ULN-L), shown in Table \ref{tab:model_specs}, which target different design points along the size/accuracy curve. All submodels use two hash functions per filter - we found that using one hash function was not always sufficient to avoid excessive collisions, while more than two hash functions increased hardware cost with no clear benefit to accuracy. All submodels within a model use the same thermometer encoding, and all models were pruned at a 30\% ratio. As shown in the table, the ensembles consisted of between 3 and 6 submodels, with ensemble accuracies ranging from 96.20\% to 98.46\%. Some of the individual submodels have accuracies as low as 80\%; however, as discussed in Section \ref{sec:ensemble}, ensembles are able to combine weak classifiers to produce accurate predictions.

\begin{table}[htbp]
\centering
\caption{Details of the selected \wname~models. SMx refers to individual sub models comprising the ensemble model.}
\begin{adjustbox}{center}
\resizebox{1.05\columnwidth}{!}{
\setlength\tabcolsep{2.5pt}
\begin{tabular}{|c|c|c|c|c|c|c|} 
 \hline
 \rowcolor{LightBlue} \centering
 Model & Sub- & Bits & Inputs & Entries & Size & Test \\
 \rowcolor{LightBlue} \centering
 & Model & /Inp & /Filter & /Filter & (KiB) & Acc.\%\\
 \hline
 %\multirow{5}{*}{\rotatebox[origin=c]{90}{ULN-S}} & Ensemble & 2 & \{\} & \{\} & 16.9 & 96.20 \\
 \cellcolor{PaleYellow} & Ensemble & 2 & \{\} & \{\} & 16.9 & 96.20 \\
 \cline{2-7}
 \cellcolor{PaleYellow} & SM0 & 2 & 12 & 64 & 7.19 & 92.91 \\
 \cellcolor{PaleYellow} & SM1 & 2 & 16 & 64 & 5.39 & 90.25 \\
 \cellcolor{PaleYellow} \multirow{-4}{*}{ULN-S} & SM2 & 2 & 20 & 64 & 4.38 & 86.16 \\
 %\hline
 \hline
 \cellcolor{PaleYellow} & Ensemble & 3 & \{\} & \{\} & 101 & 97.79 \\
 \cline{2-7}
 \cellcolor{PaleYellow} & SM0 & 3 & 12 & 64 & 10.9 & 83.54 \\
 \cellcolor{PaleYellow} & SM1 & 3 & 16 & 128 & 16.0 & 90.93 \\
 \cellcolor{PaleYellow} & SM2 & 3 & 20 & 256 & 26.0 & 92.92 \\
 \cellcolor{PaleYellow} & SM3 & 3 & 28 & 256 & 18.44 & 87.05 \\
 \cellcolor{PaleYellow} \multirow{-6}{*}{ULN-M} & SM4 & 3 & 36 & 512 & 29.38 & 80.93 \\
 %\hline
 \hline
 	
 \cellcolor{PaleYellow} & Ensemble & 7 & \{\} & \{\} & 262 & 98.46 \\
   %& mble &  &  &  &  &  \\
 \cline{2-7}
 \cellcolor{PaleYellow} & SM0 & 7 & 12 & 64 & 25.0 & 88.78 \\
 \cellcolor{PaleYellow} & SM1 & 7 & 16 & 128 & 37.7 & 93.24 \\
 \cellcolor{PaleYellow} & SM2 & 7 & 20 & 128 & 30.2 & 92.44 \\
 \cellcolor{PaleYellow} & SM3 & 7 & 24 & 256 & 50.3 & 93.92 \\
 \cellcolor{PaleYellow} & SM4 & 7 & 28 & 256 & 43.1 & 90.47 \\
 \cellcolor{PaleYellow} \multirow{-7}{*}{ULN-L} & SM5 & 7 & 32 & 512 & 75.6 & 90.44 \\
 \hline
\end{tabular}
}
\end{adjustbox}
\label{tab:model_specs}
\end{table}

\subsection{Performance of \wname~enhancements}

\begin{figure}[t]
\centerline{\includegraphics[width =1.05\columnwidth]{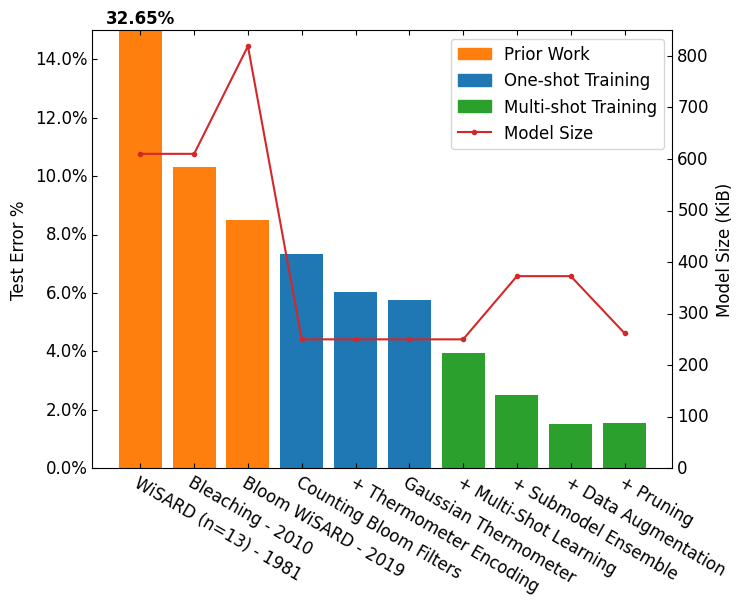}}
\vspace{-10pt}
\caption{Iterative impacts of our improvements to WNNs. We show three prior results, including the previous best model (Bloom WiSARD), and the improvements to error and model size as we sequentially incorporate \wname's optimizations. Results for MNIST dataset.}
\label{fig:improvement_breakdown}
\end{figure}

Figure \ref{fig:improvement_breakdown} illustrates the impacts of the optimizations we use in \wname. The first three models are prior work, including the original 1981 WiSARD architecture~\cite{wisard} and the state-of-the-art 2019 Bloom WiSARD architecture~\cite{Arajo2019MemoryEW}. Subsequent models incorporate progressively more of our improvements. Our final model, ULN-L, reduces error by 82\% and model size by 68\% versus Bloom WiSARD. Using multi-shot learning and ensemble models yielded significant improvements to accuracy, while pruning provides a significant reduction in model size.

\subsection{FPGA Comparison (vs FINN)}
Table \ref{tab:finn_comparison} and Figure \ref{fig:fpga_pareto} show the comparison of \wname~ with FINN. ULN-S, ULN-M, ULN-L are compared to the FINN SFC, MFC and LFC models respectively. We report accelerator latency in $\mu$s and peak throughput in thousands of inferences per second (kIPS). We also provide both the energy for a single sample in isolation (batch=1) and steady-state energy per inference (batch=$\infty$). Note that FINN does not provide some results for the MFC model.

\begin{table}[htbp]
\centering
\caption{Comparison of \wname~against FINN models SFC, MFC, LFC (FPGA). FINN rows are shaded. N/A means No Data Available.}
\begin{adjustbox}{center}
\resizebox{1.05\columnwidth}{!}{
\setlength\tabcolsep{2.5pt}
\begin{tabular}{|c|c|c|c|c|c|c|c|c|}
 \hline
 \rowcolor{LightBlue} \centering
 Model & Latency & Xput & Power & \multicolumn{2}{c|}{$\mu$J/Inf.} & LUT & BRAM & Test \\
 \rowcolor{LightBlue} \centering
 & ($\mu$s) & (kIPS) & (W) & b=1 & b=$\infty$ & & & Acc.\% \\
 \hline
 ULN-S & 0.21 & 14,286 & 1.1 & 0.234 & 0.077 & 17,319 & 0 & 96.20 \\
 \hline
 \rowcolor{light-gray}
 SFC & 0.31 & 12,361 & 7.3 & 2.263 & 0.591 & 91,131 & 4.5 & 95.83 \\
 \hline
 \hline
 ULN-M & 0.29 & 14,286 & 3.1 & 0.887 & 0.214 & 49,445 & 0 & 97.79 \\
 \hline
 \rowcolor{light-gray}
 MFC & N/A & 6,238 & 11.3 & N/A & 1.811 & N/A & N/A & 97.69 \\
 \hline
 \hline
 ULN-L & 0.94 & 4070 & 3.4 & 3.148 & 0.826 & 123,117 & 0 & 98.46 \\
 \hline
 \rowcolor{light-gray}
 LFC & 2.44 & 1561 & 8.8 & 21.5 & 5.637 & 82,988 & 396 & 98.40 \\
 \hline
\end{tabular}
}
\end{adjustbox}
\label{tab:finn_comparison}
\end{table}

Overall, we achieve a 1.4-2.6x improvement in latency and a 1.2-2.6x improvement in throughput versus FINN, with higher accuracies in all three cases. Our improvements in energy are proportionately even stronger. We reduce energy by 6.8-9.6x for a single inference and by 6.8-8.5x for steady-state inference. These improvements clearly demonstrate that WNNs can outperform even optimized BNN implementations.

All three FINN models, as well as ULN-S and ULN-M, were implemented at 200 MHz. Due to FPGA routing congestion and consequent long wires, we were limited to implementing our large design at 85 MHz.

\begin{figure}[t]
    \centering
    \begin{adjustbox}{center}
    \setlength\tabcolsep{0.0pt}
    \begin{tabular}{cc}
    \includegraphics[width=0.57\columnwidth]{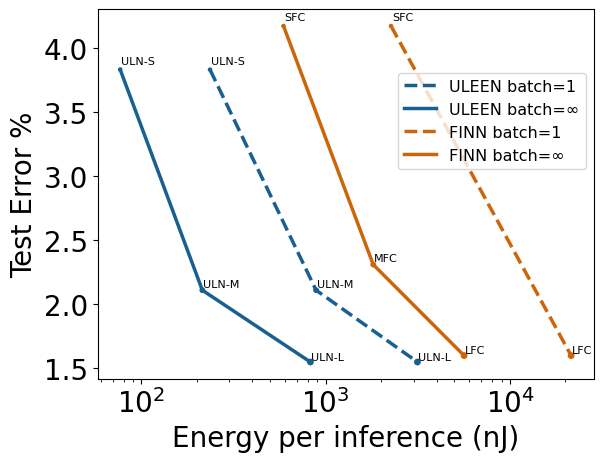} &
    \includegraphics[width=0.57\columnwidth]{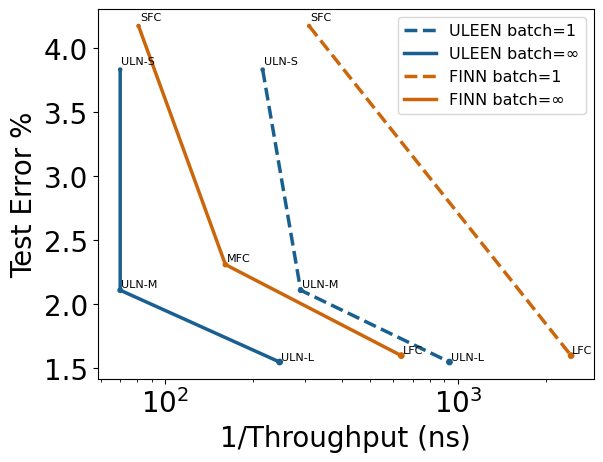} \\
    \end{tabular}
    \end{adjustbox}
    \vspace{-10pt}
    \caption{Energy and Inverse Throughput vs. Error Pareto fronts for \wname~and FINN. Both single-inference (batch=1) and steady-state (batch=$\infty$) values are shown. Note that for batch=1, inverse throughput is equal to latency.}
    \label{fig:fpga_pareto}
\end{figure}

\subsection{ASIC Comparison (vs Bit Fusion)}
\begin{figure}[htbp]
    \centerline{\includegraphics[width =0.9\columnwidth]{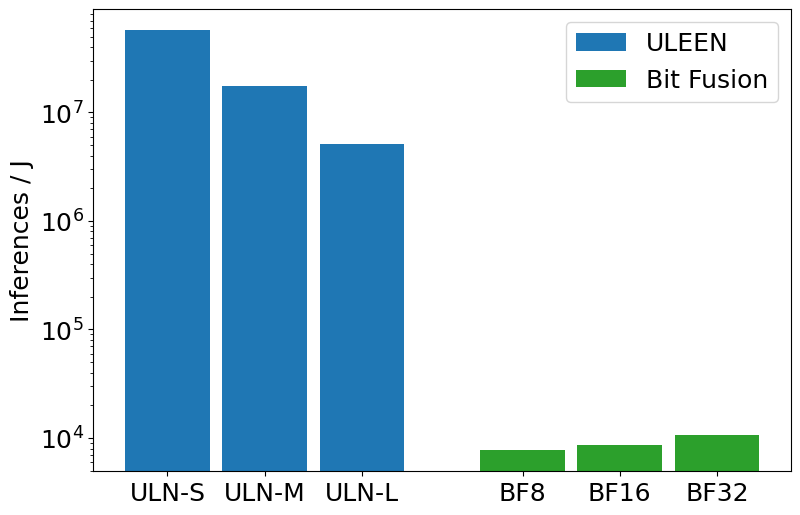}}
    \vspace{-10pt}
    \caption{Power Efficiency (Inferences per Joule) for \wname~and Bit Fusion}
    \label{fig:asic_plot}
\end{figure}

Table \ref{tab:bit_fusion_comparison} and Figure \ref{fig:asic_plot} show the comparison between Bit Fusion and \wname. Bit Fusion is designed to operate with a batch size of 16; we use the same batch size for \wname~to ensure a fair comparison. \wname~achieves a single-inference (batch=1) latency between 0.032 $\mu$s (ULN-S) and 0.056 $\mu$s (ULN-L). The fastest Bit Fusion model (BF32) requires 838 $\mu$s, but since this model performs inference in simultaneous batches of 16, a direct comparison to single-inference latency is difficult.

\begin{table}[htbp]
\centering
\caption{Comparison of \wname~against Bit Fusion/ LeNet-5 (ASIC). Bit Fusion rows are shaded.}
% Latency: 64, 84, 112
\begin{adjustbox}{center}
\resizebox{1.0\columnwidth}{!}{
\begin{tabular}{|c|c|c|c|c|c|}
 \hline
 \rowcolor{LightBlue} \centering
 Model & Xput & Power & nJ/Inf. & Area & Test \\
 \rowcolor{LightBlue} \centering
 & (kIPS) & (W) & b=16 & (mm\textsuperscript{2}) & Acc.\% \\
 \hline
 ULN-S & 55,556 & 0.84 & 17.5 & 0.61 & 96.20 \\
 \hline
 ULN-M & 55,556 & 2.58 & 57.1 & 2.09 & 97.79 \\
 \hline
 ULN-L & 38,462 & 6.23 & 195.5 & 5.22 & 98.46 \\
 \hline
 \hline
 \rowcolor{light-gray}
 BF8 & 2.0 & 0.26 & 129,731 & 0.60 & 99.35 \\
 \hline
 \rowcolor{light-gray}
 BF16 & 7.1 & 0.81 & 114,914 & 1.59 & 99.35 \\
 \hline
 \rowcolor{light-gray}
 BF32 & 19.1 & 1.79 & 93,589 & 1.65 & 99.35 \\
 \hline
\end{tabular}
}
\end{adjustbox}
\label{tab:bit_fusion_comparison}
\end{table}

Bit Fusion implements an aggressively quantized LeNet-5 model, achieving accuracy superior to ULN-L by 0.89\%. However, ULN-L uses between \textbf{479-663x} less energy and achieves \textbf{2014-19549x} greater throughput. Models such as Bit Fusion have their place in devices such as mobile phones, where an energy budget in the hundreds of $\mu$J is generally acceptable. However, \wname~presents a preferable solution for extreme edge devices, or even for larger devices if a very high throughput is needed. 

Note that the Bit Fusion accelerator is a hardware architecture intended to run any quantized convolutional neural networks. Here a quantized (ternary) LeNet-5 model is being run on Bit Fusion. Changing the size of the Bit Fusion accelerator (e.g. systolic array dimensions) impacts the performance of the accelerator, but not the accuracy of the model.
%Also, note that the Bit Fusion accelerator is a generic hardware architecture on which any neural network can be executed. Here a quantized LeNet-5 model is being run on Bit Fusion. Changing the size of the Bit Fusion accelerator (e.g. the systolic array dimensions) does not have any impact on the accuracy, which is a property of the network executed, not of the accelerator.
On the other hand, with \wname, the accelerator and model sizes are linked. \wname~establishes an interplay between accuracy, efficiency, and area, which can be explored depending on the application.
%But for \wname, the accuracy varies with the model size (accelerator size). With \wname, an interplay between accuracy and efficiency can be exercised depending on the application.

\begin{comment}
\begin{figure}[htbp]
    \centerline{\includegraphics[width =0.9\columnwidth]{figures/asic_large.png}}
    \vspace{-10pt}
    \caption{Area Efficiency (Inferences per Second per mm\textsuperscript{2}) and Power Efficiency (Inferences per J) for \wname~and Bit Fusion}
    \label{fig:asic_plot}
\end{figure}
\end{comment}

\begin{comment}
\begin{figure}[t]
    \centering
    \begin{adjustbox}{center}
    \setlength\tabcolsep{0.0pt}
    \begin{tabular}{cc}
    \includegraphics[width=0.57\columnwidth]{figures/asic_ipj.png} &
    \includegraphics[width=0.57\columnwidth]{figures/asic_eff.png} \\
    \end{tabular}
    \end{adjustbox}
    \vspace{-10pt}
    \caption{Inferences per Joule and Area Efficiency (Inferences per Second per mm\textsuperscript{2}) for \wname~and Bit Fusion}
    \label{fig:asic_plot}
\end{figure}
\end{comment}

\subsection{Model Comparison (vs Bloom WiSARD)}
Bloom WiSARD achieved accuracy comparable to large traditional WNNs at a fraction of the model size by using (binary) Bloom filters for compression~\cite{Arajo2019MemoryEW}, establishing the state of the art for efficient WNNs. We evaluate \wname~on the same nine multi-class classification datasets used for the evaluation of Bloom WiSARD, shown in Table \ref{tab:bloom_wisard_comparison}.
On MNIST, we improve test accuracy from 91.5\% (the previous best result on a sub-1MB WNN model) to 98.46\%, achieved with ULN-L.
\wname~produces a more accurate and smaller model in all cases, with a mean 46.1\% decrease in model size and 49.8\% reduction in test error.%Results are shown in Table \ref{tab:bloom_wisard_comparison} and Figure \ref{fig:bloom_wisard_comparison}.

\wname~achieves a surprising \textasciitilde99\% reduction in error on the Shuttle dataset. Shuttle is an anomaly-detection dataset in which 80\% of the training data belongs to the ``normal'' class~\cite{shuttle_dataset}. We suspect that, since Bloom WiSARD does not incorporate bleaching, the discriminator corresponding to this class became saturated during training.

\begin{table}[htbp]
\vspace{-10pt}
\centering
\caption{Accuracy and model size of \wname~(ULN-L)~vs. best prior WNN work (Bloom WiSARD)~\cite{Arajo2019MemoryEW}.}
\begin{adjustbox}{center}
\setlength\tabcolsep{2.5pt}
\begin{tabular}{|P{2.0cm}|P{1.425cm}|P{1.4cm}|P{1.425cm}|P{1.4cm}|} 
 \hline
 \rowcolor{LightBlue}
 \textbf{Model Name} & \textbf{Test Acc.\%} & \textbf{Test Acc.\%} & \textbf{Size (KiB)} & \textbf{Size (KiB)} \\
  \rowcolor{LightBlue}
   & \textbf{(Bloom} & \textbf{(\wname)} & \textbf{(Bloom} & \textbf{(\wname)} \\
    \rowcolor{LightBlue}
    & \textbf{WiSARD)} & \textbf{} & \textbf{WiSARD)} & \textbf{} \\
 \hline
 MNIST~\cite{lecun-mnisthandwrittendigit-2010}& 91.5 & 98.5 & 819 & 262\\
 \hline
 Ecoli~\cite{ecoli_dataset} & 79.9 & 87.5 & 3.28 & 0.875\\
 \hline
 Iris~\cite{iris_dataset} & 97.6 & 98.0 & 0.703 & 0.281\\
 \hline
 Letter~\cite{letter_dataset} & 84.8 & 90.0 & 91.3 & 78.0\\
 \hline
 Satimage~\cite{satimage_dataset} & 85.1 & 88.0 & 12.7 & 9.00\\
 \hline
 Shuttle~\cite{shuttle_dataset} & 86.8 & 99.9 & 3.69 & 2.63\\
 \hline
 Vehicle~\cite{vehicle_dataset} & 66.2 & 76.2 & 4.22 & 2.25\\
 \hline
 Vowel~\cite{vowel_dataset} & 87.6 & 90.0 & 6.44 & 3.44\\
 \hline
 Wine~\cite{wine_dataset} & 92.6 & 98.3 & 2.28 & 0.422\\
 \hline
\end{tabular}
\end{adjustbox}
\label{tab:bloom_wisard_comparison}
\vspace{-10pt}
\end{table}

\begin{comment}
\begin{figure*}[t]
\vspace{-3pt}
\centerline{\includegraphics[width =\textwidth]{figures/bloom_wisard_comparison.png}}
\vspace{-7pt}
\caption{\revision{The} relative errors and model sizes of the models shown in Table \ref{tab:model_specs} versus \revision{Bloom WiSARD}~\cite{Arajo2019MemoryEW}. \wname~outperforms the prior work on all nine datasets in both accuracy and model size. For the MNIST dataset, ULN-M was used for comparison.}
\label{fig:bloom_wisard_comparison}
\vspace{-10pt}
\end{figure*}
\end{comment}

\subsection{Sensitivity Analysis}

\subsubsection{Pruning}
\begin{figure}[t]
\centerline{\includegraphics[width =0.9\columnwidth]{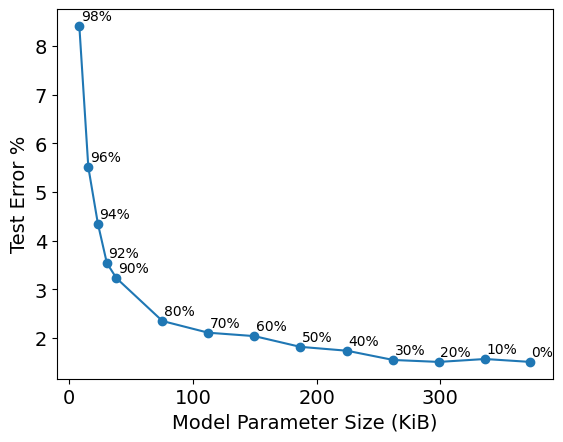}}
\vspace{-10pt}
\caption{Pruned Size vs. Error in ULN-L. The \% value next to each point denotes the amount of pruning done.}
\label{fig:pruning}
\vspace{-10pt}
\end{figure}

Figure \ref{fig:pruning} shows the model sizes and accuracies achieved by pruning ULN-L at a range of target ratios. We swept pruning in 10\% increments from 0-90\%, and in 2\% increments from 90\% to 98\%. We observe almost no loss in accuracy for pruning ratios up to 30\%, and only gradual degradation up to 80\%, which gives 97.65\% accuracy at a 75.31 KiB model size. Accuracy deteriorates rapidly past this point, but even with 98\% pruning, our test error is under 10\%.

While model size decreases proportionately with the pruning ratio, the amount of hashing required does not, since different filters may be pruned from different discriminators. Therefore, in practice it is likely less efficient to use a high pruning ratio than it is to just train a smaller model.

\begin{comment}
\subsubsection{Ensemble Aggregation}
\begin{figure}[t]
\centerline{\includegraphics[width =1.0\columnwidth]{figures/aggregation.png}}
\vspace{-10pt}
\caption{Tradeoffs in Ensemble Aggregation: Confidence vs. Summation in ULN-L. 
Using Summation provides more accuracy.
}
\label{fig:aggregation}
\end{figure}

When we discussed ensembles in Section \ref{sec:proposal}, we noted that prior work\cite{ensemble_wisard} used a confidence-based aggregation function which found the submodel with the largest difference between its strongest and next-strongest responses. \wname~instead sums the outputs of all submodels. Figure \ref{fig:aggregation} shows the result of training ULN-L using the confidence-based aggregation function, which reduces test accuracy to 97.85\%. We speculate that it is difficult for submodels to work cooperatively with confidence-based aggregation, since the output of just one submodel is propagated.
Calculating confidence scores in hardware requires additional functional units to calculate the top 2 responses of each submodel. Therefore, summation is both more accurate and simpler to implement in hardware.
\end{comment}

\subsubsection{One-shot Model Sweeping}

\begin{figure}[htbp]
    \centering
    \centerline{\includegraphics[width =0.9\columnwidth]{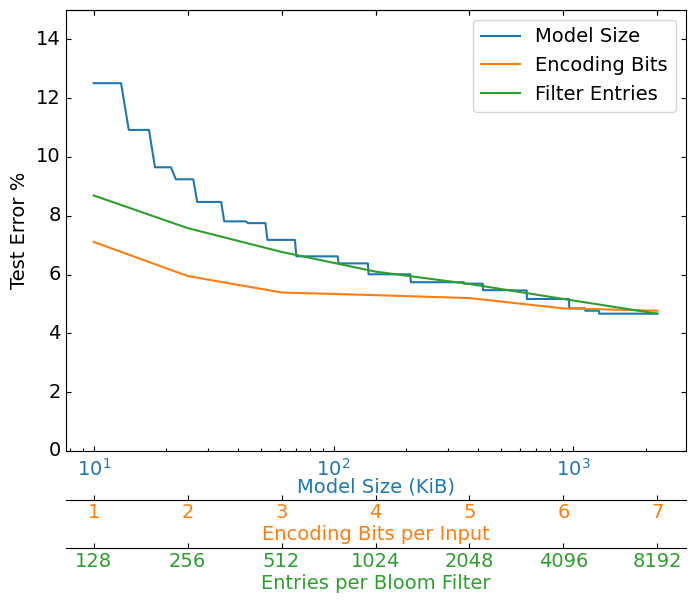}}
    \vspace{-10pt}
    \caption{Results from our one-shot sweeping runs, showing the most accurate model which was achieved with a given maximum size, the most accurate model which was achieved with a given number of encoding bits, and the most accurate model which was achieved with a given number of entries per Bloom filter.}
    \label{fig:sweep_results}
\end{figure}

Figure \ref{fig:sweep_results} summarizes the results from sweeping across a variety of model configurations using the one-shot training rule. We observe diminishing returns from increasing the number of thermometer encoding bits per Bloom filter and entries per Bloom filter (note that the entries per filter in the figure is on a log scale).

Accuracy for models trained with the one-shot technique increases roughly logarithmically with model size. We are unable to reach 96\% accuracy with a model size under 2 MB. This demonstrates the limitations of the one-shot algorithm and the importance of incorporating feedback; we are able to reach higher accuracies with much smaller models using the multi-shot learning rule.
\section{Related Work}
\label{sec:related_work}
%Prior work\cite{wisard_fpga} presented an FPGA based WiSARD model but the design presented in this paper has many enhancements such as the counting Bloom filters, inexpensive H3 hash functions and the Gaussian-based non-linear thermometer encoding.

% Prior work with nonlinear thermometers
% Other models- non-deterministic models?

%Based on XNOR Net and Bit Fusion

Our work on WNNs is suitable for edge inference applications with very low power budgets due to \wname's high energy efficiency. In this section, we discuss other works that target similar use cases.

\textbf{Quantized DNNs:}
Quantization involves transforming higher precision floating-point numbers to lower precision floating-point, fixed-point or integer~\cite{quantization_integer} values. Quantizing networks leads to reduced memory access costs and increases compute efficiency.
DNN accelerators commonly provide hardware support for lower precisions, such as 8-bit integer in Google TPU~\cite{google_tpu} or 12-bit floating-point in Microsoft Brainwave~\cite{ms_brainwave}.
Bit Fusion~\cite{bitfusion:isca:2018}, which we compare to, proposes an accelerator architecture that dynamically composes low-bitwidth compute units to match the precision requirements of DNNs.
%\wname~accelerator is significantly more efficient than a quantized Bit Fusion with the 

%\textbf{Ternary NNs:}
Ternary Weight Networks (TWNs)~\cite{TWN, DBLP:conf/iclr/ZhuHMD17, Lin2016NeuralNW} have been proposed, where weights are constrained to +1, 0 and -1. The accuracy of TWNs has been shown to be only slightly worse than the full precision networks but significantly higher than BNNs. We compare the accuracy and efficiency of our ULN-L model against a Bit Fusion accelerator running a ternarized LeNet-5. More ASIC and FPGA implementations of ternary neural networks for MNIST are presented in \cite{ternary_ijcnn}, but these have lower accuracy, throughput and energy-efficiency than our \wname~models.

\textbf{Binary NNs:}
Many methods to binarize weights or activation functions in DNNs have been proposed \cite{courbariaux2016binarized, MLSYS2020_riptide, binaryconnect, binary_net, xnor_net}. 
%BinaryConnect \cite{binaryconnect} binarizes the weights. Extensions such as BinaryNet \cite{binary_net} and XNOR-Net \cite{xnor_net} have both weights and activation functions as binary. Most of the multiplications are eliminated in both forward and backward passes, thereby achieving significant benefits in hardware implementations.
%The method of operation of WNNs differs significantly from BNNs. 
%Most BNNs are based around popcounts
Most modern BNN approaches first take the XNOR of the input with a learned weight vector, followed by a popcount operation.
Several works have explored ASIC and FPGA accelerators for BNNs (e.g., FINN~\cite{finn}, FP-BNN~\cite{fp_bnn}, and YodaNN~\cite{yodann}). As illustrated in the comparison with FINN, \wname~WNNs can achieve better latency and energy than BNNs. ReBNet \cite{rebnet} is another FPGA-based BNN accelerator for MNIST. Our throughput and accuracy are higher than ReBNet, but no energy numbers are provided in the paper.

\textbf{Compressed DNNs:}
To efficiently run DNNs on devices with limited hardware resources, model compression is used~\cite{adaptive_weight_compression, compression_block_circulant, on_demand_compression, scalpel}.
%Pruning involves removing unimportant parameters (e.g. Deep Compression \cite{deep_compression}) and also reducing the number of parameters by replacing larger filters with smaller filters or decreasing the number of channels in a layer (e.g., SqueezeNet \cite{squeezenet}). This involves retraining networks to avoid accuracy degradation. Pruning is commonly used for edge devices \cite{patdnn}.
Pruning~\cite{deep_compression, squeezenet, patdnn} reduces the total number of parameters, but may require retraining to avoid accuracy degradation.
Weight sparsity in CNNs has been used to reduce computation complexity~\cite{cnvlutin, scnn, cambricon_x, cambricon_s}, and bit sparsity is exploited in~\cite{bit_pragmatic, diffy}.
%However, WNNs can be trained in a one-shot process: a binarized input is broken into $n$-tuples, each $n$-tuple addressing a RAM position that will be incremented every time is accessed during the training procedure. 
We use two forms of compression in \wname: Bloom filters to reduce the size of the RAM nodes, and pruning to reduce the number of RAM nodes.

\textbf{Prior WNNs:}
%Early WNNs~\cite{} and more recent designs~\cite{} report
%Prior work introduced concepts of bleaching \cite{Grieco:2010:PPE:1751674.1751890} however, we are the first to introduce bleaching in a hardware implementable WNN model.
Prior work~\cite{Arajo2019MemoryEW, santiago2020weightless} demonstrated that replacing the RAM nodes in WiSARD with Bloom filters improved memory efficiency and model robustness, though we are the first to use counting Bloom filters (which facilitates bleaching). We are also the first to implement a Bloom filter-based WNN in hardware, as prior work used an impractical hashing algorithm.
%In discriminator-based WNNs, like WiSARD and ULEEN, each input pattern is broken into $n$-tuples, each $n$-tuple addressing a memory position. Although the binary system is a quasi-default, other numeric bases could be directly used in memory addressing. 
In~\cite{6855847}, a WiSARD-based rock-paper-scissors player was built using a ternary system, where each input digit value was uniquely associated to \textit{rock}, \textit{paper}, or \textit{scissors} states.
%Prior work ~\cite{Arajo2019MemoryEW}\cite{santiago2020weightless}, used a double-hashing technique based on the MurmurHash\cite{MurmurHash} algorithm. However, this approach requires many arithmetic operations (e.g., 5 multiplications to hash a 32-bit value), and is therefore impractical in hardware. We identified an arithmetic-free approach based on universal families of hash functions which is much less expensive to implement. Thus, while prior work presented ideas of software-implemented Bloom filters, our design incorporates realistic filters which abide by hardware constraints. 
Prior memory-efficient WNNs~\cite{Grieco:2010:PPE:1751674.1751890,Arajo2019MemoryEW} report up to 91.5\% accuracy on MNIST while our improvements raise accuracy to 98.46\%. 

%we suggest changing the term "symbolic" to "symbolist" so that the paragraph would not be mistaken as to referring to algebraic computation. 

\textbf{Symbolist Approaches:}
%Hardware development for more complex symbolist processing, such as logic and functional programming, that flourish in the nineties, has lost ground to simpler symbolist paradigms such as decision trees and Bayesian inference. In fact, symbolist processing generally works well for small-scale problems and leads to small-footprint, energy-efficient architectures. However, these models typically require sophisticated, heavy domain-dependent preprocessing, or incur in large (e.g., half a million LUTs ~\cite{decisiontree}) or very customized circuit implementations (e.g., clockless probabilistic p-bits ~\cite{10.3389/fncom.2021.584797}).
Symbolist paradigms such as decision trees and Bayesian inference generally work well for small-scale problems.
%and lead to small-footprint, energy-efficient architectures. 
However, these models typically require sophisticated, heavy domain-dependent preprocessing, or incur large (e.g., half a million LUTs~\cite{decisiontree}) or very customized circuit implementations (e.g., clockless probabilistic p-bits~\cite{10.3389/fncom.2021.584797}).

%Symbolism generally works for small-scale problems and leads to small-footprint, energy-efficient architectures. However, symbolic models, such as decision trees, typically require sophisticated preprocessing, domain knowledge, or large models. Decisions trees and other symbolic approaches ~\cite{decisiontree}, symbolic need heavy domain-dependent preprocessing.  (eg: half a million LUTs~\cite{decisiontree}).

\textbf{Other edge DNN acceleration approaches:}
Microcontroller-based approaches to edge inference, such as TinyML, have attracted a great deal of interest recently due to their ability to use inexpensive off-the-shelf hardware~\cite{arduino}. However, these approaches to machine learning are thousands of times slower than dedicated accelerators. A TinyML MNIST solution on Arduino Nano using downscaled 8x8 images had >130,000x lower throughput and more than 32,000x higher latency than our FPGA implementation of ULN-M on 28x28 images, in addition to being less accurate~\cite{arduino}.
%MLPerf tiny solutions ....report
\section{Conclusion}
\label{sec:conclusion}

%While most machine learning research centers around DNNs, we explore an alternate neural model, the Weightless Neural Network, for edge inference. We incorporate enhancements such as counting Bloom filters, inexpensive H3 hash functions and a Gaussian-based non-linear thermometer encoding into the WiSARD weightless neural model\revision{, improving state-of-the-art WNN MNIST accuracy~\cite{Arajo2019MemoryEW} from 91.5\% to 95.2\%}. The proposed \wname~ architecture is compared to state-of-the-art weightless models as well as MLPs and CNNs of similar accuracy. An FPGA accelerator for \wname~ is also presented. Compared to prior WNNs, \wname~ reduces error by 41\% and model size by 51\% across nine datasets. Compared to \revision{iso-accuracy} MLP models, \wname~ consumes \textasciitilde20\% of the total energy while reducing latency by \textasciitilde85\%. \revision{Energy/latency improvements over CNNs are even larger, although CNNs have higher accuracy.}
Inference on the extreme edge demands new approaches to machine learning. While existing techniques use quantized DNNs or BNNs, we propose \wname, an approach based on Weightless Neural Networks. \wname~introduces and incorporates counting Bloom filters, Gaussian thermometer encoding, additive ensembles, pruning, and a novel multi-shot training algorithm to achieve superior energy and latency compared to quantized DNNs and BNNs.
%reduce model size by 46.1\% and 49.8\% versus the prior WNN state of the art. 
We present FPGA and ASIC implementations of an inference accelerator for \wname. Compared against the FINN FPGA-based platform for BNNs, we improve latency by 1.4-2.6x, throughput by 1.2-2.6x, and 
%steady-state
energy by 6.8-8.5x. Compared to the Bit Fusion low-precision DNN ASIC, we reduce energy by 479-663x and throughput by 2014-19549x, though with slightly lower accuracy. The energy per inference for the most accurate FINN model is 5.637 $\mu$J whereas \wname~ consumes only 0.826 $\mu$J per inference for the largest model presented. For the Bit Fusion model, the energy per inference ranges from 93.5-129 $\mu$J, whereas  \wname~ models only consume 0.01-0.2 $\mu$J (i.e. more than 400x improvement), illustrating the potential of \wname~ for extreme edge devices.

The most direct opportunity for future work that we see is the development of convolutional WNNs. Convolution is important for good accuracy on larger image datasets such as ImageNet. Although support for convolution adds model and hardware complexity, we believe that WNNs have the potential to excel in this domain, as they can learn nonlinear filters, something that is not possible with traditional CNNs.

Our results show that WNNs hold substantial promise for inference on the edge. WNNs are efficient, but have historically trailed BNNs and quantized DNNs in accuracy; the improvements in \wname~narrow or close that gap, demonstrating their suitability for ultra-low-energy and high-throughput inference.

\textit{\textbf{Acknowledgement:} This research was supported in part by Semiconductor Research Corporation (SRC) Task 3015.001/3016.001 and National Science Foundation grant number 1763848. Any opinions, findings, conclusions or recommendations are those of the authors and not of the funding agencies.}

%%
%% The next two lines define the bibliography style to be used, and
%% the bibliography file.
\bibliographystyle{IEEEtran}
\bibliography{bibliography}

% Generated by IEEEtran.bst, version: 1.14 (2015/08/26)
\begin{thebibliography}{10}
\providecommand{\url}[1]{#1}
\csname url@samestyle\endcsname
\providecommand{\newblock}{\relax}
\providecommand{\bibinfo}[2]{#2}
\providecommand{\BIBentrySTDinterwordspacing}{\spaceskip=0pt\relax}
\providecommand{\BIBentryALTinterwordstretchfactor}{4}
\providecommand{\BIBentryALTinterwordspacing}{\spaceskip=\fontdimen2\font plus
\BIBentryALTinterwordstretchfactor\fontdimen3\font minus
  \fontdimen4\font\relax}
\providecommand{\BIBforeignlanguage}[2]{{%
\expandafter\ifx\csname l@#1\endcsname\relax
\typeout{** WARNING: IEEEtran.bst: No hyphenation pattern has been}%
\typeout{** loaded for the language `#1'. Using the pattern for}%
\typeout{** the default language instead.}%
\else
\language=\csname l@#1\endcsname
\fi
#2}}
\providecommand{\BIBdecl}{\relax}
\BIBdecl

\bibitem{russa2015}
O.~Russakovsky, J.~Deng, H.~Su, J.~Krause, S.~Satheesh, S.~Ma, Z.~Huang,
  A.~Karpathy, A.~Khosla, M.~Bernstein, A.~C. Berg, and L.~Fei-Fei, ``Imagenet
  large scale visual recognition challenge,'' \emph{International Journal of
  Computer Vision}, vol. 115, pp. 211--252, 4 2015.

\bibitem{gholami2021survey}
A.~Gholami, S.~Kim, Z.~Dong, Z.~Yao, M.~W. Mahoney, and K.~Keutzer, ``A survey
  of quantization methods for efficient neural network inference,'' 2021.

\bibitem{once_for_all}
\BIBentryALTinterwordspacing
H.~Cai, C.~Gan, T.~Wang, Z.~Zhang, and S.~Han, ``Once-for-all: Train one
  network and specialize it for efficient deployment,'' 2019. [Online].
  Available: \url{https://arxiv.org/abs/1908.09791}
\BIBentrySTDinterwordspacing

\bibitem{extreme_edge_sensors}
\BIBentryALTinterwordspacing
E.~Covi, E.~Donati, X.~Liang, D.~Kappel, H.~Heidari, M.~Payvand, and W.~Wang,
  ``Adaptive extreme edge computing for wearable devices,'' \emph{Frontiers in
  Neuroscience}, vol.~15, 2021. [Online]. Available:
  \url{https://www.frontiersin.org/article/10.3389/fnins.2021.611300}
\BIBentrySTDinterwordspacing

\bibitem{extreme_edge_overview}
J.~Portilla, G.~Mujica, J.-S. Lee, and T.~Riesgo, ``The extreme edge at the
  bottom of the internet of things: A review,'' \emph{IEEE Sensors Journal},
  vol.~19, no.~9, pp. 3179--3190, 2019.

\bibitem{binaryconnect}
M.~Courbariaux, Y.~Bengio, and J.-P. David, ``Binaryconnect: Training deep
  neural networks with binary weights during propagations,'' in
  \emph{Proceedings of the 28th International Conference on Neural Information
  Processing Systems - Volume 2}, ser. NIPS'15.\hskip 1em plus 0.5em minus
  0.4em\relax Cambridge, MA, USA: MIT Press, 2015, p. 3123–3131.

\bibitem{binary_net}
\BIBentryALTinterwordspacing
``Binarized neural networks,'' \emph{CoRR}, vol. abs/1602.02505, 2016,
  withdrawn. [Online]. Available: \url{http://arxiv.org/abs/1602.02505}
\BIBentrySTDinterwordspacing

\bibitem{finn}
\BIBentryALTinterwordspacing
Y.~Umuroglu, N.~J. Fraser, G.~Gambardella, M.~Blott, P.~Leong, M.~Jahre, and
  K.~Vissers, ``Finn: A framework for fast, scalable binarized neural network
  inference,'' in \emph{Proceedings of the 2017 ACM/SIGDA International
  Symposium on Field-Programmable Gate Arrays}, ser. FPGA '17.\hskip 1em plus
  0.5em minus 0.4em\relax New York, NY, USA: Association for Computing
  Machinery, 2017, p. 65–74. [Online]. Available:
  \url{https://doi.org/10.1145/3020078.3021744}
\BIBentrySTDinterwordspacing

\bibitem{xnor_net}
\BIBentryALTinterwordspacing
M.~Rastegari, V.~Ordonez, J.~Redmon, and A.~Farhadi, ``Xnor-net: Imagenet
  classification using binary convolutional neural networks,'' \emph{CoRR},
  vol. abs/1603.05279, 2016. [Online]. Available:
  \url{http://arxiv.org/abs/1603.05279}
\BIBentrySTDinterwordspacing

\bibitem{courbariaux2016binarized}
M.~Courbariaux, I.~Hubara, D.~Soudry, R.~El-Yaniv, and Y.~Bengio, ``Binarized
  neural networks: Training deep neural networks with weights and activations
  constrained to +1 or -1,'' 2016.

\bibitem{wnn_intro_esann}
I.~Aleksander, M.~De~Gregorio, F.~França, P.~Lima, and H.~Morton, ``A brief
  introduction to weightless neural systems,'' in \emph{17th European Symposium
  on Artificial Neural Networks (ESANN)}, 04 2009, pp. 299--305.

\bibitem{weightless_review}
T.~Ludermir, A.~de~Carvalho, A.~Braga, and M.~Souto, ``Weightless neural
  models: A review of current and past works,'' \emph{Neural Computing
  Surveys}, vol.~2, pp. 41--61, 01 1999.

\bibitem{straight_through_estimator}
\BIBentryALTinterwordspacing
P.~Yin, J.~Lyu, S.~Zhang, S.~J. Osher, Y.~Qi, and J.~Xin, ``Understanding
  straight-through estimator in training activation quantized neural nets,''
  \emph{CoRR}, vol. abs/1903.05662, 2019. [Online]. Available:
  \url{http://arxiv.org/abs/1903.05662}
\BIBentrySTDinterwordspacing

\bibitem{Grieco:2010:PPE:1751674.1751890}
\BIBentryALTinterwordspacing
B.~P.~A. Grieco, P.~M.~V. Lima, M.~De~Gregorio, and F.~M.~G. Fran\c{c}a,
  ``Producing pattern examples from "mental" images,'' \emph{Neurocomput.},
  vol.~73, no. 7-9, pp. 1057--1064, Mar. 2010. [Online]. Available:
  \url{http://dx.doi.org/10.1016/j.neucom.2009.11.015}
\BIBentrySTDinterwordspacing

\bibitem{bleaching}
D.~Carvalho, H.~Carneiro, F.~França, and P.~Lima, ``B-bleaching : Agile
  overtraining avoidance in the wisard weightless neural classifier,'' in
  \emph{ESANN}, 04 2013.

\bibitem{bitfusion:isca:2018}
H.~Sharma, J.~Park, N.~Suda, L.~Lai, B.~Chau, V.~Chandra, and H.~Esmaeilzadeh,
  ``Bit fusion: Bit-level dynamically composable architecture for accelerating
  deep neural network,'' in \emph{45th {ACM/IEEE} Annual International
  Symposium on Computer Architecture, {ISCA} 2018, Los Angeles, CA, USA, June
  1-6, 2018}, 2018.

\bibitem{10.5555/284803.284804}
R.~Rohwer and M.~Morciniec, ``The theoretical and experimental status of the
  $n$-tuple classifier,'' \emph{Neural Netw.}, vol.~11, no.~1, p. 1–14, Jan.
  1998.

\bibitem{unsupervised_wnn}
I.~Wickert and F.~França, ``Validating an unsupervised weightless
  perceptron,'' in \emph{Proceedings of the 9th International Conference on
  Neural Information Processing, 2002. ICONIP '02.}, 12 2002, pp. 537 -- 541
  vol.2.

\bibitem{cluswisard}
D.~O. Cardoso, D.~Carvalho, D.~S.~F. Alves, D.~F.~P. de~Souza, H.~C.~C.
  Carneiro, C.~E. Pedreira, P.~M.~V. Lima, and F.~M.~G. Fran\c{c}a, ``Financial
  credit analysis via a clustering weightless neural classifier,''
  \emph{Neurocomputing}, vol. 183, pp. 70--78, 2016.

\bibitem{wisard}
\BIBentryALTinterwordspacing
I.~Aleksander, W.~Thomas, and P.~Bowden, ``{WISARD·a radical step forward in
  image recognition},'' \emph{Sensor Review}, vol.~4, no.~3, pp. 120--124,
  1984. [Online]. Available:
  \url{https://www.emerald.com/insight/content/doi/10.1108/eb007637/full/html}
\BIBentrySTDinterwordspacing

\bibitem{vc_dimension}
V.~Vapnik and A.~Chervonenkis, \emph{On the uniform convergence of relative
  frequencies of events to their probabilities}.\hskip 1em plus 0.5em minus
  0.4em\relax Springer, 01 2015, pp. 11--30.

\bibitem{wisard_vc}
H.~Carneiro, C.~Pedreira, F.~França, and P.~Lima, ``The exact vc dimension of
  the wisard n-tuple classifier,'' \emph{Neural Computation}, pp. 1--32, 11
  2018.

\bibitem{Arajo2019MemoryEW}
L.~S. de~Ara{\'u}jo, L.~D. Verona, F.~M. Rangel, F.~F. de~Faria, D.~S.
  Menasch{\'e}, W.~Caarls, M.~Breternitz, S.~Kundu, P.~M.~V. Lima, and F.~M.~G.
  França, ``Memory efficient weightless neural network using bloom filter,''
  in \emph{ESANN}, 2019.

\bibitem{CARTER1979143}
\BIBentryALTinterwordspacing
J.~Carter and M.~N. Wegman, ``Universal classes of hash functions,''
  \emph{Journal of Computer and System Sciences}, vol.~18, no.~2, pp. 143--154,
  1979. [Online]. Available:
  \url{https://www.sciencedirect.com/science/article/pii/0022000079900448}
\BIBentrySTDinterwordspacing

\bibitem{MurmurHash}
A.~Appleby, ``Murmurhash3,'' \url{https://github.com/aappleby/smhasher}, 2016.

\bibitem{wisard_encoding}
A.~Kappaun, K.~Camargo, F.~Rangel, F.~Firmino, P.~M.~V. Lima, and J.~Oliveira,
  ``Evaluating binary encoding techniques for wisard,'' in \emph{2016 5th
  Brazilian Conference on Intelligent Systems (BRACIS)}, 2016, pp. 103--108.

\bibitem{Xavier2020DetectionOE}
P.~Xavier, M.~D. Gregorio, F.~M.~G. França, and P.~M.~V. Lima, ``Detection of
  elementary particles with the wisard n-tuple classifier,'' in \emph{ESANN},
  2020.

\bibitem{ensemble_review}
T.~G. Dietterich, ``Ensemble methods in machine learning,'' in \emph{Multiple
  Classifier Systems}.\hskip 1em plus 0.5em minus 0.4em\relax Berlin,
  Heidelberg: Springer Berlin Heidelberg, 2000, pp. 1--15.

\bibitem{ensemble_wisard}
\BIBentryALTinterwordspacing
L.~A. {Lusquino Filho}, L.~F. Oliveira, A.~L. Filho, G.~P. Guarisa, L.~M.
  Felix, P.~M. Lima, and F.~M. França, ``Extending the weightless wisard
  classifier for regression,'' \emph{Neurocomputing}, vol. 416, pp. 280--291,
  2020. [Online]. Available:
  \url{https://www.sciencedirect.com/science/article/pii/S092523122030504X}
\BIBentrySTDinterwordspacing

\bibitem{deep_compression}
S.~Han, H.~Mao, and W.~Dally, ``Deep compression: Compressing deep neural
  networks with pruning, trained quantization and huffman coding,'' in
  \emph{International Conference on Learning Representations (ICLR)}, 10 2016.

\bibitem{scalpel}
J.~Yu, A.~Lukefahr, D.~Palframan, G.~Dasika, R.~Das, and S.~Mahlke, ``Scalpel:
  Customizing dnn pruning to the underlying hardware parallelism,'' in
  \emph{2017 ACM/IEEE 44th Annual International Symposium on Computer
  Architecture (ISCA)}, 2017, pp. 548--560.

\bibitem{dnn_weight_pruning_framework}
T.~Zhang, S.~Ye, K.~Zhang, J.~Tang, W.~Wen, M.~Fardad, and Y.~Wang, ``A
  systematic dnn weight pruning framework using alternating direction method of
  multipliers,'' in \emph{Computer Vision -- ECCV 2018}, V.~Ferrari, M.~Hebert,
  C.~Sminchisescu, and Y.~Weiss, Eds.\hskip 1em plus 0.5em minus 0.4em\relax
  Cham: Springer International Publishing, 2018, pp. 191--207.

\bibitem{patdnn}
\BIBentryALTinterwordspacing
W.~Niu, X.~Ma, S.~Lin, S.~Wang, X.~Qian, X.~Lin, Y.~Wang, and B.~Ren,
  \emph{PatDNN: Achieving Real-Time DNN Execution on Mobile Devices with
  Pattern-Based Weight Pruning}.\hskip 1em plus 0.5em minus 0.4em\relax New
  York, NY, USA: Association for Computing Machinery, 2020, p. 907–922.
  [Online]. Available: \url{https://doi.org/10.1145/3373376.3378534}
\BIBentrySTDinterwordspacing

\bibitem{adam}
D.~Kingma and J.~Ba, ``Adam: A method for stochastic optimization,''
  \emph{International Conference on Learning Representations}, 12 2014.

\bibitem{dropout}
N.~Srivastava, G.~Hinton, A.~Krizhevsky, I.~Sutskever, and R.~Salakhutdinov,
  ``Dropout: A simple way to prevent neural networks from overfitting,''
  \emph{J. Mach. Learn. Res.}, vol.~15, no.~1, p. 1929–1958, jan 2014.

\bibitem{lecun-mnisthandwrittendigit-2010}
\BIBentryALTinterwordspacing
Y.~LeCun and C.~Cortes, ``{MNIST} handwritten digit database,''
  http://yann.lecun.com/exdb/mnist/, 2010. [Online]. Available:
  \url{http://yann.lecun.com/exdb/mnist/}
\BIBentrySTDinterwordspacing

\bibitem{finn_documentation}
\BIBentryALTinterwordspacing
Xilinx. (2021) {FINN: Dataflow compiler for QNN inference on FPGAs}. [Online].
  Available: \url{https://xilinx.github.io/finn/}
\BIBentrySTDinterwordspacing

\bibitem{bit_fusion_documentation}
\BIBentryALTinterwordspacing
G.~I. o.~T. ACT~Lab. (2018) {Bit Fusion: Bit-Level Dynamically Composable
  Architecture for Accelerating DNNs}. [Online]. Available:
  \url{http://www.act-lab.org/artifacts/bitfusion/}
\BIBentrySTDinterwordspacing

\bibitem{TWN}
\BIBentryALTinterwordspacing
F.~Li and B.~Liu, ``Ternary weight networks,'' \emph{CoRR}, vol.
  abs/1605.04711, 2016. [Online]. Available:
  \url{http://arxiv.org/abs/1605.04711}
\BIBentrySTDinterwordspacing

\bibitem{freepdk45}
\BIBentryALTinterwordspacing
NCSU. (2018) {FreePDK45}. [Online]. Available:
  \url{https://www.eda.ncsu.edu/wiki/FreePDK45:Contents}
\BIBentrySTDinterwordspacing

\bibitem{cacti}
\BIBentryALTinterwordspacing
H.~Packard. (2017) {Cacti}. [Online]. Available:
  \url{https://github.com/HewlettPackard/cacti/}
\BIBentrySTDinterwordspacing

\bibitem{mako}
\BIBentryALTinterwordspacing
M.~Bayer, ``Mako templates for python,'' 2021. [Online]. Available:
  \url{https://www.makotemplates.org/}
\BIBentrySTDinterwordspacing

\bibitem{shuttle_dataset}
\BIBentryALTinterwordspacing
J.~Catlett, ``Statlog (shuttle) data set.'' [Online]. Available:
  \url{https://archive.ics.uci.edu/ml/datasets/Statlog+(Shuttle)}
\BIBentrySTDinterwordspacing

\bibitem{ecoli_dataset}
\BIBentryALTinterwordspacing
K.~Nakai, ``Ecoli data set.'' [Online]. Available:
  \url{https://archive.ics.uci.edu/ml/datasets/ecoli}
\BIBentrySTDinterwordspacing

\bibitem{iris_dataset}
\BIBentryALTinterwordspacing
R.~Fisher, ``Iris data set.'' [Online]. Available:
  \url{https://archive.ics.uci.edu/ml/datasets/iris}
\BIBentrySTDinterwordspacing

\bibitem{letter_dataset}
\BIBentryALTinterwordspacing
D.~J. Slate, ``Letter recognition data set.'' [Online]. Available:
  \url{https://archive.ics.uci.edu/ml/datasets/letter+recognition}
\BIBentrySTDinterwordspacing

\bibitem{satimage_dataset}
\BIBentryALTinterwordspacing
A.~Srinivasan, ``Statlog (landsat satellite) data set.'' [Online]. Available:
  \url{https://archive.ics.uci.edu/ml/datasets/Statlog+(Landsat+Satellite)}
\BIBentrySTDinterwordspacing

\bibitem{vehicle_dataset}
\BIBentryALTinterwordspacing
P.~Mowforth and B.~Shepherd, ``Statlog (vehicle silhouettes) data set.''
  [Online]. Available:
  \url{https://archive.ics.uci.edu/ml/datasets/Statlog+(Vehicle+Silhouettes)}
\BIBentrySTDinterwordspacing

\bibitem{vowel_dataset}
\BIBentryALTinterwordspacing
D.~Deterding, M.~Niranjan, and T.~Robinson, ``Connectionist bench (vowel
  recognition - deterding data) data set.'' [Online]. Available:
  \url{https://archive.ics.uci.edu/ml/datasets/Connectionist+Bench+(Vowel+Recognition+-+Deterding+Data)}
\BIBentrySTDinterwordspacing

\bibitem{wine_dataset}
\BIBentryALTinterwordspacing
S.~Aeberhard, ``Wine data set.'' [Online]. Available:
  \url{https://archive.ics.uci.edu/ml/datasets/wine}
\BIBentrySTDinterwordspacing

\bibitem{quantization_integer}
\BIBentryALTinterwordspacing
B.~Jacob, S.~Kligys, B.~Chen, M.~Zhu, M.~Tang, A.~Howard, H.~Adam, and
  D.~Kalenichenko, ``Quantization and training of neural networks for efficient
  integer-arithmetic-only inference,'' 2017. [Online]. Available:
  \url{https://arxiv.org/abs/1712.05877}
\BIBentrySTDinterwordspacing

\bibitem{google_tpu}
\BIBentryALTinterwordspacing
N.~P. Jouppi, C.~Young, N.~Patil, D.~Patterson, G.~Agrawal, R.~Bajwa, S.~Bates,
  S.~Bhatia, N.~Boden, A.~Borchers, R.~Boyle, P.~luc Cantin, C.~Chao, C.~Clark,
  J.~Coriell, M.~Daley, M.~Dau, J.~Dean, B.~Gelb, T.~V. Ghaemmaghami,
  R.~Gottipati, W.~Gulland, R.~Hagmann, C.~R. Ho, D.~Hogberg, J.~Hu, R.~Hundt,
  D.~Hurt, J.~Ibarz, A.~Jaffey, A.~Jaworski, A.~Kaplan, H.~Khaitan, A.~Koch,
  N.~Kumar, S.~Lacy, J.~Laudon, J.~Law, D.~Le, C.~Leary, Z.~Liu, K.~Lucke,
  A.~Lundin, G.~MacKean, A.~Maggiore, M.~Mahony, K.~Miller, R.~Nagarajan,
  R.~Narayanaswami, R.~Ni, K.~Nix, T.~Norrie, M.~Omernick, N.~Penukonda,
  A.~Phelps, and J.~Ross, ``In-datacenter performance analysis of a tensor
  processing unit,'' 2017. [Online]. Available:
  \url{https://arxiv.org/pdf/1704.04760.pdf}
\BIBentrySTDinterwordspacing

\bibitem{ms_brainwave}
J.~{Fowers}, K.~{Ovtcharov}, M.~{Papamichael}, T.~{Massengill}, M.~{Liu},
  D.~{Lo}, S.~{Alkalay}, M.~{Haselman}, L.~{Adams}, M.~{Ghandi}, S.~{Heil},
  P.~{Patel}, A.~{Sapek}, G.~{Weisz}, L.~{Woods}, S.~{Lanka}, S.~K.
  {Reinhardt}, A.~M. {Caulfield}, E.~S. {Chung}, and D.~{Burger}, ``{A
  Configurable Cloud-Scale DNN Processor for Real-Time AI},'' in \emph{2018
  ACM/IEEE 45th Annual International Symposium on Computer Architecture
  (ISCA)}, 2018, pp. 1--14.

\bibitem{DBLP:conf/iclr/ZhuHMD17}
\BIBentryALTinterwordspacing
C.~Zhu, S.~Han, H.~Mao, and W.~J. Dally, ``Trained ternary quantization,'' in
  \emph{5th International Conference on Learning Representations, {ICLR} 2017,
  Toulon, France, April 24-26, 2017, Conference Track Proceedings}.\hskip 1em
  plus 0.5em minus 0.4em\relax OpenReview.net, 2017. [Online]. Available:
  \url{https://openreview.net/forum?id=S1\_pAu9xl}
\BIBentrySTDinterwordspacing

\bibitem{Lin2016NeuralNW}
Z.~Lin, M.~Courbariaux, R.~Memisevic, and Y.~Bengio, ``Neural networks with few
  multiplications,'' \emph{CoRR}, vol. abs/1510.03009, 2016.

\bibitem{ternary_ijcnn}
H.~Alemdar, V.~Leroy, A.~Prost-Boucle, and F.~Pétrot, ``Ternary neural
  networks for resource-efficient ai applications,'' in \emph{2017
  International Joint Conference on Neural Networks (IJCNN)}, 2017, pp.
  2547--2554.

\bibitem{MLSYS2020_riptide}
\BIBentryALTinterwordspacing
J.~Fromm, M.~Cowan, M.~Philipose, L.~Ceze, and S.~Patel, ``Riptide: Fast
  end-to-end binarized neural networks,'' in \emph{Proceedings of Machine
  Learning and Systems}, I.~Dhillon, D.~Papailiopoulos, and V.~Sze, Eds.,
  vol.~2, 2020, pp. 379--389. [Online]. Available:
  \url{https://proceedings.mlsys.org/paper/2020/file/2a79ea27c279e471f4d180b08d62b00a-Paper.pdf}
\BIBentrySTDinterwordspacing

\bibitem{fp_bnn}
\BIBentryALTinterwordspacing
S.~Liang, S.~Yin, L.~Liu, W.~Luk, and S.~Wei, ``Fp-bnn: Binarized neural
  network on fpga,'' \emph{Neurocomputing}, vol. 275, pp. 1072--1086, 2018.
  [Online]. Available:
  \url{https://www.sciencedirect.com/science/article/pii/S0925231217315655}
\BIBentrySTDinterwordspacing

\bibitem{yodann}
R.~Andri, L.~Cavigelli, D.~Rossi, and L.~Benini, ``Yodann: An architecture for
  ultralow power binary-weight cnn acceleration,'' \emph{IEEE Transactions on
  Computer-Aided Design of Integrated Circuits and Systems}, vol.~37, no.~1,
  pp. 48--60, 2018.

\bibitem{rebnet}
M.~Ghasemzadeh, M.~Samragh, and F.~Koushanfar, ``Rebnet: Residual binarized
  neural network,'' in \emph{2018 IEEE 26th Annual International Symposium on
  Field-Programmable Custom Computing Machines (FCCM)}, 2018, pp. 57--64.

\bibitem{adaptive_weight_compression}
J.~H. Ko, D.~Kim, T.~Na, J.~Kung, and S.~Mukhopadhyay, ``Adaptive weight
  compression for memory-efficient neural networks,'' in \emph{Design,
  Automation Test in Europe Conference Exhibition (DATE), 2017}, 2017, pp.
  199--204.

\bibitem{compression_block_circulant}
C.~Ding, S.~Liao, Y.~Wang, Z.~Li, N.~Liu, Y.~Zhuo, C.~Wang, X.~Qian, Y.~Bai,
  G.~Yuan, X.~Ma, Y.~Zhang, J.~Tang, Q.~Qiu, X.~Lin, and B.~Yuan, ``Circnn:
  Accelerating and compressing deep neural networks using block-circulant
  weight matrices,'' in \emph{2017 50th Annual IEEE/ACM International Symposium
  on Microarchitecture (MICRO)}, 2017, pp. 395--408.

\bibitem{on_demand_compression}
\BIBentryALTinterwordspacing
S.~Liu, Y.~Lin, Z.~Zhou, K.~Nan, H.~Liu, and J.~Du, ``On-demand deep model
  compression for mobile devices: A usage-driven model selection framework,''
  in \emph{Proceedings of the 16th Annual International Conference on Mobile
  Systems, Applications, and Services}, ser. MobiSys '18.\hskip 1em plus 0.5em
  minus 0.4em\relax New York, NY, USA: Association for Computing Machinery,
  2018, p. 389–400. [Online]. Available:
  \url{https://doi.org/10.1145/3210240.3210337}
\BIBentrySTDinterwordspacing

\bibitem{squeezenet}
F.~N. Iandola, M.~W. Moskewicz, K.~Ashraf, S.~Han, W.~J. Dally, and K.~Keutzer,
  ``Squeezenet: Alexnet-level accuracy with 50x fewer parameters and <1mb model
  size,'' \emph{ArXiv}, vol. abs/1602.07360, 2016.

\bibitem{cnvlutin}
J.~Albericio, P.~Judd, T.~Hetherington, T.~Aamodt, N.~E. Jerger, and
  A.~Moshovos, ``Cnvlutin: Ineffectual-neuron-free deep neural network
  computing,'' in \emph{2016 ACM/IEEE 43rd Annual International Symposium on
  Computer Architecture (ISCA)}, 2016, pp. 1--13.

\bibitem{scnn}
A.~Parashar, M.~Rhu, A.~Mukkara, A.~Puglielli, R.~Venkatesan, B.~Khailany,
  J.~Emer, S.~W. Keckler, and W.~J. Dally, ``Scnn: An accelerator for
  compressed-sparse convolutional neural networks,'' in \emph{2017 ACM/IEEE
  44th Annual International Symposium on Computer Architecture (ISCA)}, 2017,
  pp. 27--40.

\bibitem{cambricon_x}
S.~Zhang, Z.~Du, L.~Zhang, H.~Lan, S.~Liu, L.~Li, Q.~Guo, T.~Chen, and Y.~Chen,
  ``Cambricon-x: An accelerator for sparse neural networks,'' in \emph{2016
  49th Annual IEEE/ACM International Symposium on Microarchitecture (MICRO)},
  2016, pp. 1--12.

\bibitem{cambricon_s}
X.~Zhou, Z.~Du, Q.~Guo, S.~Liu, C.~Liu, C.~Wang, X.~Zhou, L.~Li, T.~Chen, and
  Y.~Chen, ``Cambricon-s: Addressing irregularity in sparse neural networks
  through a cooperative software/hardware approach,'' in \emph{2018 51st Annual
  IEEE/ACM International Symposium on Microarchitecture (MICRO)}, 2018, pp.
  15--28.

\bibitem{bit_pragmatic}
J.~Albericio, A.~Delmás, P.~Judd, S.~Sharify, G.~O’Leary, R.~Genov, and
  A.~Moshovos, ``Bit-pragmatic deep neural network computing,'' in \emph{2017
  50th Annual IEEE/ACM International Symposium on Microarchitecture (MICRO)},
  2017, pp. 382--394.

\bibitem{diffy}
M.~Mahmoud, K.~Siu, and A.~Moshovos, ``Diffy: a déjà vu-free differential
  deep neural network accelerator,'' in \emph{2018 51st Annual IEEE/ACM
  International Symposium on Microarchitecture (MICRO)}, 2018, pp. 134--147.

\bibitem{santiago2020weightless}
L.~Santiago, L.~Verona, F.~Rangel, F.~Firmino, D.~S. Menasch{\'e}, W.~Caarls,
  M.~Breternitz~Jr, S.~Kundu, P.~M. Lima, and F.~M. Fran{\c{c}}a, ``Weightless
  neural networks as memory segmented bloom filters,'' \emph{Neurocomputing},
  vol. 416, pp. 292--304, 2020.

\bibitem{6855847}
D.~F.~d. Souza, H.~C. Carneiro, F.~M. França, and P.~M. Lima,
  ``Rock-paper-scissors wisard,'' in \emph{2013 BRICS Congress on Computational
  Intelligence and 11th Brazilian Congress on Computational Intelligence},
  2013, pp. 178--182.

\bibitem{decisiontree}
R.~Choudhury, S.~R. Ahamed, and P.~Guha, ``Efficient hardware implementation of
  decision tree training accelerator,'' in
  \emph{https://ieeexplore.ieee.org/document/942607345}.

\bibitem{10.3389/fncom.2021.584797}
\BIBentryALTinterwordspacing
R.~Faria, J.~Kaiser, K.~Y. Camsari, and S.~Datta, ``Hardware design for
  autonomous bayesian networks,'' \emph{Frontiers in Computational
  Neuroscience}, vol.~15, 2021. [Online]. Available:
  \url{https://www.frontiersin.org/article/10.3389/fncom.2021.584797}
\BIBentrySTDinterwordspacing

\bibitem{arduino}
\BIBentryALTinterwordspacing
Simone, ``{TinyML or Arduino and STM32: Convolutional Neural Network (CNN)
  Example, Accessed Nov 22, 2021},'' 2020. [Online]. Available:
  \url{https://eloquentarduino.github.io/2020/11/tinyml-on-arduino-and-stm32-cnn-convolutional-neural-network-example/}
\BIBentrySTDinterwordspacing

\end{thebibliography}

\end{document}